\documentclass[11pt]{article}

\usepackage{graphics,graphicx}
\usepackage{amsmath,amsthm,amssymb}
\usepackage[spanish, english]{babel}
\usepackage{textcomp}
\usepackage{rotating}
\usepackage{lscape}
\usepackage{longtable}
\usepackage{color}
\usepackage[latin1,utf8]{inputenc}
\usepackage[T1]{fontenc}
\usepackage{fancyhdr}
\usepackage{pst-all}
\usepackage[bookmarks = true, colorlinks=true, linkcolor = blue, citecolor = blue, menucolor = black, urlcolor = black, plainpages=false]{hyperref}
\usepackage[T1]{fontenc}
\usepackage{rawfonts}
\usepackage{amsfonts}
\usepackage{enumerate}
\usepackage{pstricks}
\usepackage{layout}
\usepackage{colortbl}
\usepackage{multimedia}
\usepackage{times}
\usepackage{array}
\usepackage{supertabular}
\usepackage{xspace}
\usepackage{ragged2e}
\usepackage{multicol}
\usepackage[active]{srcltx}
\usepackage{makeidx}
\usepackage{subcaption}
\usepackage{dsfont}
\usepackage{bbm}
\usepackage{epstopdf}
\usepackage{breakcites}

\newtheorem{Theorem}{Theorem}[section]
\newtheorem{Definition}[Theorem]{Definition}

\newtheorem{Proposition}[]{Proposition}

\newtheorem{Prop}{Proposition}

\newenvironment{Proof}[1][Proof]{{\text{#1.\;\;}}}{\hfill $\square$}%
\numberwithin{equation}{section} 
\numberwithin{Theorem}{section}

\begin{document}




\title{\Large{Directional Multivariate Extremes in Environmental Phenomena}}


\author{Raúl Torres (ratorres@est-econ.uc3m.es) \\
Carlo De Michele  (carlo.demichele@polimi.it) \\
Henry Laniado (hlaniado@est-econ.uc3m.es) \\
Rosa E. Lillo (lillo@est-econ.uc3m.es)}
\date{}
\maketitle

\begin{abstract}

Several environmental phenomena can be described by different correlated variables that must be considered jointly in order to be more representative of the nature of these phenomena. For such events, identification of extremes is inappropriate if it is based on marginal analysis. Extremes have usually been linked to the notion of quantile, which is an important tool to analyze risk in the univariate setting. We propose to identify multivariate extremes and analyze environmental phenomena in terms of the directional multivariate quantile, which allows us to analyze the data considering all the variables implied in the phenomena, as well as look at the data in interesting directions that can better describe an environmental catastrophe. Since there are many references in the literature that propose extremes detection based on copula models, we also generalize the copula method by introducing the directional approach. Advantages and disadvantages of the non-parametric proposal that we introduce and the copula methods are provided in the paper. We show with simulated and real data sets how by considering the first principal component direction we can improve the visualization of extremes. Finally, two cases of study are analyzed: a synthetic case of flood risk at a dam (a $3-$variable case), and a real case study of sea storms (a $5-$variable case).
\end{abstract}

%
%


\section{Introduction}
Serious economic and social consequences are generally associated with extreme environmental events such as floods, storms and droughts (\cite{chebana2}), which are usually defined in terms of several correlated variables.
For instance, rainfall is characterized by storm intensity and duration (e.g. \cite{demichele0,sydm}); air quality is described in terms of levels of ozone and nitrogen dioxide (e.g. \cite{chebana1,heffernan}; floods are modeled by their peak, volume and duration (e.g. \cite{shiau,demichele1,grimaldi1,chebana2}); droughts are modeled by volume, duration and magnitude (e.g. \cite{kim,demichele3,sydm2015}) and sea storms are represented by wave height, peak, direction and duration (e.g. \cite{demichele2}). Consequently, extremes detection cannot be made on the basis of a univariate analysis.

There are references in the literature that tackle multivariate extreme detection. Some studies use copulas since the work by \cite{demichele0}, for example \cite{salvadori,setal,setal1,setal2}, whom also define multivariate versions of the return period\footnote{For further information of the return period we refer to \cite{nature}}, and \cite{grimaldi2}. Another alternative is given by \cite{chebana1} through depth functions. However, both alternatives have drawbacks when they have to be implemented in high dimensional scenarios. Copulas due to their intrinsic parametric nature are difficult to estimate in large dimensions, and depth functions are problematic due to the lack of computational implementation in most of the cases. Therefore, the first contribution of the paper is to introduce a method to detect extremes based on a non-parametric procedure suitable for high dimensional analysis.

On the other hand, extremes have been traditionally analyzed in one dimension by considering only the probabilities of exceeding quantiles related to either the distribution function or the survival function. In other words, observations are considered extreme if they are associated to lower values or upper values of the variable, which is equivalent to looking at the data in one of the two possible directions $\{-1,+1\}$. Some extensions of quantiles to the bivariate case have been proposed in \cite{fpsll,shiau,salvadori,ep} and to the $n-$dimensional setting in \cite{gupta,setal,pateiro,bernardino,bernardino3}. The generalizations of the quantile notion in all the previous references consider, as in the univariate case, the directions associated with the distribution function or the survival function.

But why not look at the data with different perspectives and take advantage of the inherent complexity of the $n-$dimensional setting the data lives in? There exist infinite directions to look at the data from a reference point that could help with the  accuracy  of the analysis and the interpretation of the results. Attempts have been  made  considering alternative directions, for instance  \cite{laniadoIme} and \cite{torres1st} developed a financial application where the risk of losses is analyzed considering the direction of the  investment weight composition in a portfolio; \cite{cascosRM1,hapasi} and \cite{mizera} have applied a directional setting to define quantile trimmings. Hence, the second contribution of this paper is to outline a general approach to detect directional multivariate extremes, which can be useful in other statistical areas apart from  environmental sciences.

The definition of directional multivariate extremes is based on the directional multivariate quantile introduced in \cite{laniadoWP} and \cite{torres1st}, where the free parameter of direction included can be chosen considering external information such as anthropogenic forces generating today's the environmental global-change (see \cite{hegerl}). Specifically, we propose to use principal component analysis (\textit{PCA}) in the environmental framework since the visualization of the extremes improves with respect to the use of the classical directions, as is shown in two cases of study. Firstly,  we use the  flood model proposed in \cite{setal} for the Ceppo Morelli dam in Italy  to perform a Monte Carlo study for a time window of 1000 years. Our approach improves previous results by the reduction of the ratio of false positives (regular observations which are classified as extremes). Secondly, we perform a study of sea storms considering variables such as wave height, water level and storm duration which characterize storm magnitude, storm direction, and inter-arrival time which provide information about the period of calm between two successive storms. The study shows relevant  differences with the work by \cite{demichele2} such as the computational feasibility of the method in the  $5-$dimensional setting and also the visualization of the  extremes with cross-sectional plots, where it is shown how the classical theory identifies an excessive number of observations as extremes.

The third contribution of the paper is to introduce the directional approach in the copula method. We obtain results that establish the equivalence between the directional approach and the copula based methods. It is also shown with the simulations across the document how using a mixture of both settings (directional and copula approach), we can describe better a multivariate system.

The structure of the paper is the following: Section 2 introduces the notion of directional multivariate extremes and the non-parametric procedure to carry out the identification in practice. In Section 3 and Section 4, we motivate the use of principal components (\textit{PCA}) to get an interesting direction of analysis in real case studies. Section 5 presents a summary of the classical methodology based on copulas, and theoretical results linking copulas and the notion of directions. We also present in this section some examples of the pros and cons for the extreme identification using our directional non-parametric procedure or the extended copula method. Finally, Section 6 presents some conclusions.

\section{Methodology}
In this section, we present a description of the procedure to identify directional multivariate extremes based on the directional setting proposed in \cite{torres1st}, a non-parametric algorithm for practical implementation and the motivation of the first \textit{PCA} direction.

\subsection{Directional Multivariate Extreme Value Analysis}

The directional multivariate setting is defined in terms of the \textit{oriented orthant} introduced in \cite{laniadoIme}.\\

An oriented orthant in $\mathbb{R}^{n}$ with vertex $\mathbf{x}$ in the direction $\mathbf{u}$ is defined by,
\begin{equation}\label{eq:convexcone}
	\mathfrak{C}_{\mathbf{x}}^{R_{\mathbf{u}}}=\{\mathbf{z}\in \mathbb{R}^{n}:R_{\mathbf{u}}(\mathbf{z}-\mathbf{x})\geq 0\}.
\end{equation}

where $\mathbf{u}\in\{\mathbf{v}\in\mathbb{R}^{n}:||\mathbf{v}||=1\}$ and $R_{\mathbf{u}}$ is an orthogonal matrix such that $R_{\mathbf{u}}\mathbf{u}=\mathbf{e}$, with $\mathbf{e}=\frac{\sqrt{n}}{n}[1,...,1]'$.\\

\begin{figure}[htbp]
\begin{center}
\includegraphics[height=30mm,width=30mm]{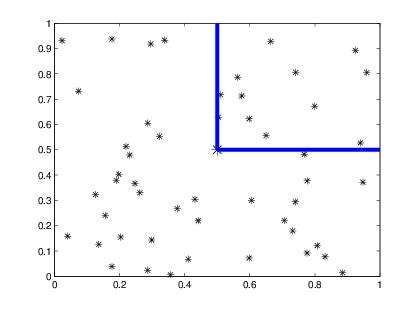}\hspace{1cm}\includegraphics[height=30mm,width=30mm]{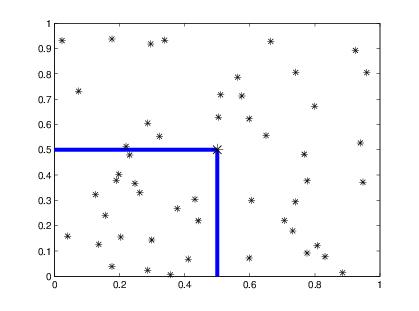}\\
\centerline{(A)\;$\mathbf{u}=\mathbf{e}$\hspace{2.5cm} (B)\;$\mathbf{u}=-\mathbf{e}$}\ \ \\
\includegraphics[height=30mm,width=30mm]{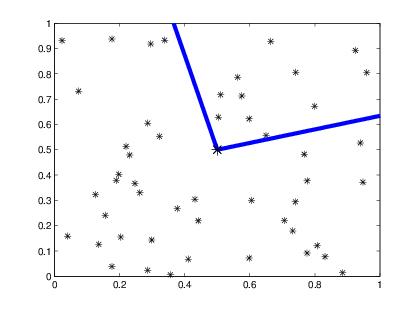}\hspace{1cm}
\includegraphics[height=30mm,width=30mm]{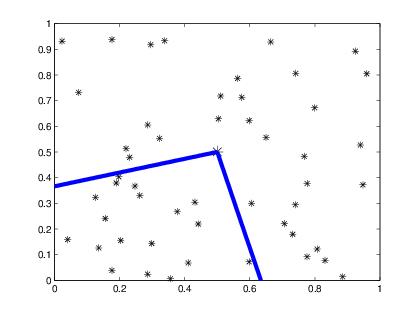}\\
\centerline{\ \ \hspace{0.3cm}(C)\;$\mathbf{u}=(1/2,\sqrt{3}/2)$ \ \ \hspace{0.5cm} (D)\;$\mathbf{u}=(-1/2,\sqrt{3}/2)$}\ \ \\
\caption{Examples of QR oriented orthants with the same vertex but different directions}\label{fig:orthants}
\vspace{-0.5cm}
\end{center}
\end{figure}

Note that an oriented orthant is a  translation and a rotation of the non-negative euclidean orthant toward  a new vertex in point $\mathbf{x}$ and a new direction $\mathbf{u}$. As is explained in \cite{torres1st},  $R_{\mathbf{u}}$ is not unique for $n\geq 3$. Then, in order to guarantee uniqueness in the orthogonal transformation, the QR oriented orthant is defined in \cite{torres1st}. The QR oriented orthant is an oriented orthant with a specific orthogonal transformation   $R_{\mathbf{u}}$  (the details about the construction of the ortogonal matrix can been seen in Appendix A.), and  hereafter, the QR oriented orthant is denoted by  $\mathfrak{C}_{\mathbf{x}}^{\mathbf{u}}$. Figure \ref{fig:orthants} shows examples of the divisions in the bivariate plane that can be performed using the concept of QR oriented orthant for different directions. Note that the direction $\mathbf{u}=\mathbf{e}$ generates the rotation matrix $R_{\mathbf{u}}$ equal to the identity matrix.

If $n=1$ (univariate setting), there are only two possible directions $\{-1,1\}$ and the corresponding orthants at vertex $x$ are the intervals  $\{(-\infty,x]$, $[x,\infty)\}$), respectively. Then, in terms of probability, they represent the valuation of the distribution and survival functions in $x$. But, when $n>1$, note that the values of the distribution and survival functions at some point $\mathbf{x}$ correspond to the probability of the QR oriented orthants with vertexes in directions $-\mathbf{e},\,\mathbf{e}$ respectively. In the multivariate extremes literature, there are many studies that use those functions as a natural way to extend different procedures from the univariate extreme analysis (e.g. \cite{sydm,demichele1,ep,bernardino3}).

However, infinite directions are possible when $n>1$, which motivates the directional approach, since  more important than using the distribution and survival functions for a random vector $\mathbf{X}$, could be using directly the probability measure of the random vector to describe the extremes properly. To clarify ideas, one can think in the bivariate setting and a random vector $\mathbf{X}$ with negative dependence. Then, it seems more convenient to use the complementary part of the division of the plane than the pair of directions $\{-\mathbf{e},\mathbf{e}\}$, i.e., to use the directions given by $\left\{\left(-1/\sqrt{2},1/\sqrt{2}\right),\left(1/\sqrt{2}, -1/\sqrt{2}\right)\right\}$ (e.g. \cite{bcsll,chebana2}), hence the importance of the directional approach. Hereafter we call \textit{classical directions} the collection of $2^{n}$ orthants that divide naturally the hyper-plane, i.e., the collection of unitary $n-$dimensional vectors with components in $\{-1,\, 1\}$. Now, we can introduce the necessary tools to attain the main purposes of our work, after motivation of the directions.\\

A directional multivariate quantile of a random vector $\mathbf{X}$ at level $\alpha$ in direction $\mathbf{u}$, is defined as,

\begin{equation}\label{dirQ}
	\mathcal{Q}_{\mathbf{X}}(\alpha,\mathbf{u})=\partial\{\mathbf{z}\in \mathbb{R}^{n}:\mathbb{P}\left[\mathfrak{C}_{\mathbf{z}}
	^{\mathbf{u}}\}\right]\leq \alpha\}
\end{equation}

where $0 < \alpha < 1$, and $\partial$ means the boundary of a set. Once a value of $\alpha$ is fixed (near to $0$ for extreme value analysis), $\mathcal{Q}_{\mathbf{X}}(\alpha,\mathbf{u})$ divides the space into two sets:

\begin{itemize}
\item \textit{The upper $\alpha-$level set in direction $\mathbf{u}$}:
\begin{equation}\label{upperSet}
		\mathcal{U}_{\mathbf{X}}(\alpha,\mathbf{u})=\{\mathbf{z}\in \mathbb{R}^{n}:\mathbb{P}\left[\mathfrak{C}_{\mathbf{z}}
		^{\mathbf{u}}\}\right]< \alpha\}
\end{equation}

\item \textit{The lower $\alpha-$level set in direction $\mathbf{u}$}:

\begin{equation}\label{lowerSet}
		\mathcal{L}_{\mathbf{X}}(\alpha,\mathbf{u})=\{\mathbf{z}\in \mathbb{R}^{n}:\mathbb{P}\left[\mathfrak{C}_{\mathbf{z}}
		^{\mathbf{u}}\}\right]> \alpha\}.
\end{equation}
\end{itemize}

These sets motivate the definition of  extreme related to the pair  $(\alpha, \mathbf{u})$ as those points exceeding the threshold given by the hyper-curve $\mathcal{Q}_{\mathbf{X}}(\alpha,\mathbf{u})$, i.e., we characterize the \textit{extreme events} as those points belonging to the associated \textit{upper level set}. The \textit{risky points} are the ones belonging to \textit{the directional multivariate quantile} $\mathcal{Q}_{\mathbf{X}}(\alpha,\mathbf{u})$ and the \textit{non-risky points} are those in the \textit{lower level set}.

Note that expressions  (\ref{dirQ}), (\ref{upperSet}) and (\ref{lowerSet}) with $\mathbf{u}=\mathbf{e}$ and values of $\alpha$ close to zero are the multivariate extension of the univariate quantile definition based on the survival function. Now, if we rewrite those expressions in terms of the pair $(1-\alpha, -\mathbf{u})$ and reversing the inequalities, we obtain the corresponding quantile extension related to the distribution function. However, these two alternatives are not equivalent for dimension $n\geq 2$ unlike the univariate case. Such duality can be also seen in the approaches AND and OR defined in \cite{demichele2}), or the UPPER and LOWER differentiation given in \cite{ep,bernardino}. But, without loss of generality, we have decided to implement the extreme detection analysis in terms of the survival analogy, because a key relationship can be established between the extremes given by (\ref{upperSet}) and those associated to the arguments $(1-\alpha,-\mathbf{u})$ reversing the inequalities (see Corollary 4.3 in \cite{torres1st}); that is,

\begin{equation}\label{eq:setsRel}
\underline{\mathcal{U}}_{\mathbf{X}}(\alpha,\mathbf{u}):=\{\mathbf{z}\in \mathbb{R}^{n}:\mathbb{P}\left[\mathfrak{C}_{\mathbf{z}}^{-\mathbf{u}}\}\right]> 1-\alpha\} \subset \mathcal{U}_{\mathbf{X}}(\alpha,\mathbf{u}),
\end{equation} 

Then, in terms of risks, relation (2.5) allows us to consider risk aversion; that is, we would expect more extreme events which corresponds to a conservative position. In the supplementary material, one can find an extensive explanation of the selected way to extend. Now, we describe a non-parametric procedure to estimate the extreme thresholds, i.e., the directional multivariate quantiles, as well as, the lower and upper level sets for a dataset.

\subsection{Non-parametric procedure and suitable direction of analysis}\label{subsec:dirApp}

As we mentioned in the Introduction, one  of the contributions of this paper is to provide a non-parametric algorithm to estimate the quantiles. It is remarkable that most of the references that deal with the multivariate extreme identification problem  are based on copula procedures that have inherent weaknesses due to the complex process of parameter estimation and the absence of computational feasibility in high dimensions. Therefore, we try to improve these issues by introducing a pseudo-algorithm based on the empirical distribution in order to get the level sets we are interested in. Firstly, we fix a preliminary notation:

\begin{itemize}
\item $\mathbf{X}_{m}:=\{\mathbf{x_{1}},\cdots,\mathbf{x_{m}}\}$,  sample data from the random vector $\mathbf{X}$,
\item $\mathbb{P}_{\mathbf{X}_{m}}[\cdot]$ is the empirical probability law of $\mathbf{X}_{m}$,
\item $\hat{\mathcal{Q}}_{\mathbf{X}_{m}}^{h}(\alpha,\mathbf{u}):=\left\lbrace \mathbf{x_{j}} : \left|\mathbb{P}_{\mathbf{X}_{m}}\left[\mathfrak{C}_{\mathbf{x_{j}}}^{\mathbf{u}}\right]-\alpha\right|\leq h\right\rbrace$ the sample quantile curve with a slack $h$, avoiding an empty set of estimated quantiles.
\item $\hat{\mathcal{U}}_{\mathbf{X}_{m}}^{h}(\alpha,\mathbf{u}):=\left\lbrace \mathbf{x_{j}} : \mathbb{P}_{\mathbf{X}_{m}}\left[\mathfrak{C}_{\mathbf{x_{j}}}^{\mathbf{u}}\right] < \alpha - h\right\rbrace$ the sample upper $\alpha-$level set with a slack $h$,
\item $\hat{\mathcal{L}}_{\mathbf{X}_{m}}^{h}(\alpha,\mathbf{u}):=\left\lbrace \mathbf{x_{j}} : \mathbb{P}_{\mathbf{X}_{m}}\left[\mathfrak{C}_{\mathbf{x_{j}}}^{\mathbf{u}}\right] > \alpha + h\right\rbrace$ the sample lowe $\alpha-$level set with a slack $h$.
\end{itemize}

Once defined the direction of analysis and the parameter $\alpha$, it is possible to estimate the directional multivariate quantile and the level sets through the following pseudo-algorithm:

\begin{quotation}
	Input: $\mathbf{u}$, $\alpha$, $h$ and the multivariate sample $\mathbf{X}_{m}$.
	
	$\text{ for } i=1 \text{ to } m$\vspace{1pt}
	
	$\quad P_{i}=\mathbb{P}_{\mathbf{X}_{m}}\left[\mathfrak{C}
	_{\mathbf{x}_{i}}^{\mathbf{u}}\right]$,\vspace{1pt}
	
	$\quad \text{If } |P_{i}-\alpha|\leq h$\vspace{1pt}
	
	$\qquad \mathbf{x}_{i}\in \hat{\mathcal{Q}}_{\mathbf{X}_{m}}^{h}(\alpha,\mathbf{u})$,\vspace{1pt}
	
	$\quad$ end\vspace{1pt}
	
	$\quad \text{If } P_{i} < \alpha - h$\vspace{1pt}
	
	$\qquad \mathbf{x}_{i}\in \hat{\mathcal{U}}_{\mathbf{X}_{m}}^{h}(\alpha,\mathbf{u})$,\vspace{1pt}
	
	$\quad$ end\vspace{1pt}
	
	$\quad \text{If } P_{i} > \alpha + h$\vspace{1pt}
	
	$\qquad \mathbf{x}_{i}\in \hat{\mathcal{L}}_{\mathbf{X}_{m}}^{h}(\alpha,\mathbf{u})$,\vspace{1pt}
	
	$\quad$ end\vspace{1pt}

\end{quotation}

\begin{figure}[htbp]
\begin{center}
\includegraphics[height=30mm,width=30mm]{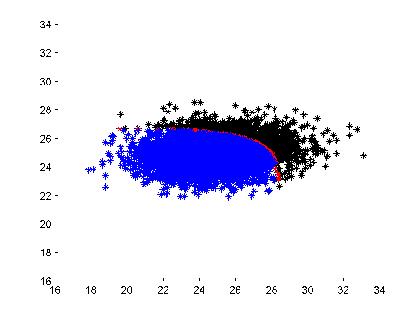}\hspace{0.5cm}\includegraphics[height=30mm,width=30mm]{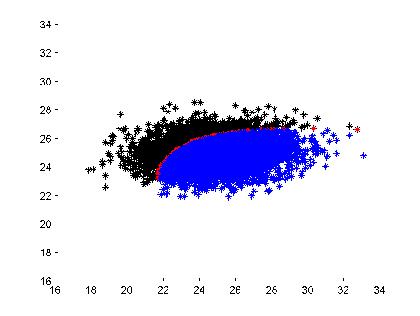}\hspace{0.5cm}\includegraphics[height=30mm,width=30mm]{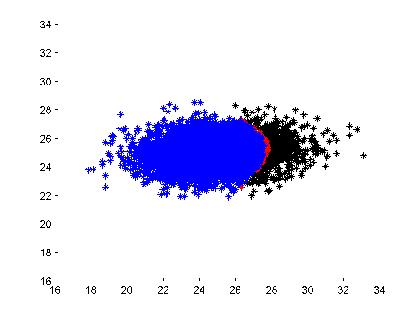}
\centerline{\ \ \hspace{1cm}(A)\;$\mathbf{u}=\mathbf{e}$\hspace{1.2cm} (B)\;$\mathbf{u}=\left(-\frac{1}{\sqrt{2}}, \frac{1}{\sqrt{2}}\right)$\hspace{0.5cm} (C)\;$\mathbf{u}=$ First \textit{PCA}}\ \ \\%
\caption{Directional Extremes at $\alpha=5\%$}\label{fig:levelSet}
\vspace{-0.5cm}
\end{center}
\end{figure}

As output, we get an estimation of the directional $\alpha-$quantile, or in other words, the hyper surface of thresholds in the selected direction of analysis, $\hat{\mathcal{Q}}_{\mathbf{X}_{m}}^{h}(\alpha,\mathbf{u})$. We also obtain those points belonging to the non-risky level set, $\hat{\mathcal{L}}_{\mathbf{X}_{m}}^{h}(\alpha,\mathbf{u})$, and the extreme level set, $\hat{\mathcal{U}}_{\mathbf{X}_{m}}^{h}(\alpha,\mathbf{u})$. An example  is presented in Figure \ref{fig:levelSet}, where simulated data from a bivariate normal distribution with $\boldsymbol\mu = [25,25]$, $\sigma_{1}^{2} = 4$, $\sigma_{2}^{2}=1$, $\rho_{1,2}=0.15$  is considered. We show the three sets of observations, the directional $5\%-$quantiles in red, the \textit{upper $5\%-$level set} or \textit{extreme level-set} in black and the \textit{lower $5\%-$level set} or \textit{non-risky level set} in blue. We have used  three different directions: the classical direction $\mathbf{e}$ (survival distribution), the complementary bivariate direction $(-1/\sqrt{2}, 1/\sqrt{2})$ and the direction given by the first \textit{PCA}. It can be observed how the identification of extremes varies according to changes in the direction in which you look at the data, for the same level $\alpha=5\%$.

Notice that different contexts or phenomena could lead to consider different particular directions of interest. For instance in portfolio theory, the direction given by the portfolio weights of investments is of particular interest because it takes into account the losses due to the composition of the investment in the portfolio (see \cite{laniadoIme,torres1st}). On the other hand, researchers in environmental science could consider important other directions more related to the phenomenon of analysis. In any case, we want to motivate here an interesting way to obtain a relevant direction of analysis by considering the principal component analysis (\textit{PCA}), which is an important statistical multivariate tool that describes the information about variability of the data jointly considered.
It is well known that the first component  provides the direction that accumulates the maximum amount of uncertainty of the data by the strongest linear combination representing the behavior of the system. We have tested this direction as a good candidate for the analysis in the following two cases of study.

\section{Case study: flood risk at a dam}

\cite{setal} presented a $3-$dimensional model to describe floods occurring at Ceppo Morelli dam, located in Piedmont region, north-western Italy. In that work, the following three variables are considered: maximum annual \textit{Peak} (\textit{Q} in $m^3/s$), maximum annual \textit{Volume} (\textit{V} in $10^6 m^3$) and initial \textit{Water Level} (\textit{L} in $m$) in the reservoir before the flood event. The model that links all the variables was estimated using a copula approach to capture the correlation structure and generalized extreme value distributions (\textit{GEV}) to describe the marginal behavior of $Q$ and $V$, while a non-parametric Normal kernel for $L$. However, for simplicity in the calculations, the simulation has been made using GEV for all the marginals. Then, the model was finally completed through Sklar's theorem and nested copula procedures, (see Appendix B. for an overview of the tools). Figure \ref{fig:ceppo} shows the scatterplot and the $3D-$plot of the dataset used in \cite{setal}.

\begin{figure}[htbp]
\begin{center}
\includegraphics[height=5cm,width=5cm]{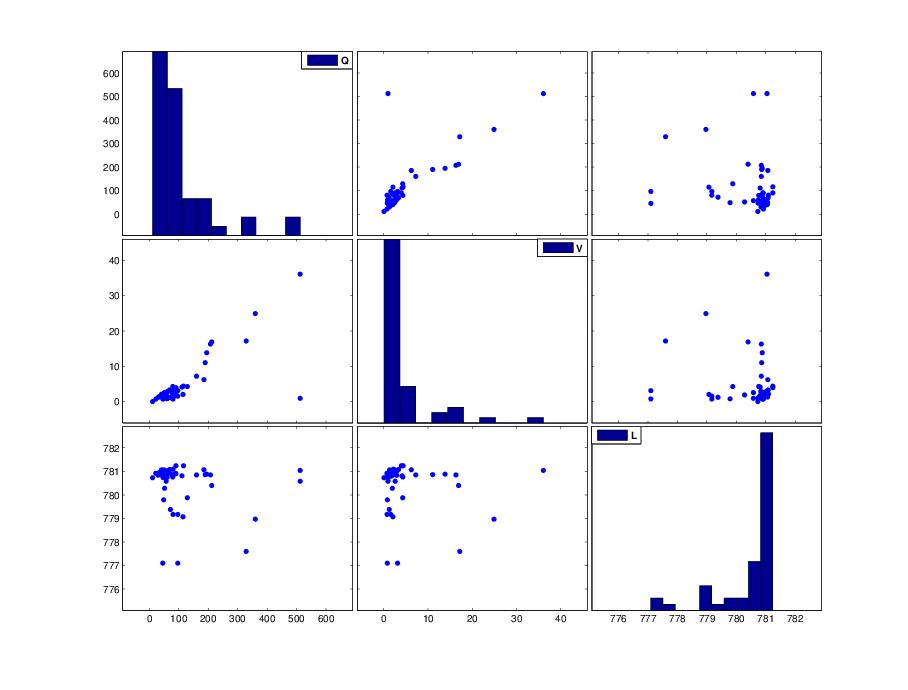}\hspace{0.5cm}\includegraphics[height=5cm,width=5cm]{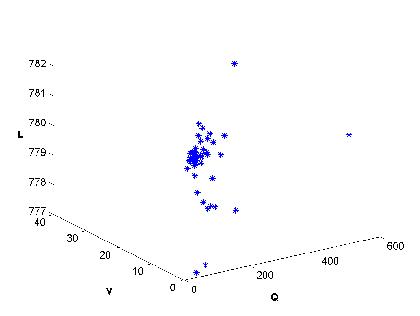}
\caption{Cross-sections and $3D-$plot of the dataset from Ceppo Morelli dam}\label{fig:ceppo}
\vspace{-0.5cm}
\end{center}
\end{figure}

The specifications of the model are given in Table \ref{table:ceppo}, the (\textit{GEV}) distributions fitted for each variable with the corresponding parameters of location $\epsilon$, scale $\beta$ and shape $\gamma$ and  the copula model to recover the joint distribution of $\mathbf{X}=(Q,V,L)$.  The pair $(Q,V)$ has associated  a Gumbel copula with positive dependence, the pairs, $(Q,L)$ and $(V,L)$ are modeled using the copula product. Finally,  the flood copula model is given by $C_{QVL}$ after a nesting procedure.

\begin{table}[htbp]
\begin{center}
\small{\begin{tabular}{||c|c|c||}\hline\hline
$Q$ & $v_{1}=F_{Q}(q)$ & \textit{GEV} with $\epsilon=59.358 m^3/s,\, \beta=36.203 m^3/s,\, \gamma=0.368$\\ \hline
$V$ & $v_{2}=F_{V}(v)$ & \textit{GEV} with $\epsilon=1.7231 m^3,\ \beta=1.5246 m^3,\, \gamma=0.6149$\\ \hline
$L$ & $v_{3}=F_{L}(l)$ & \textit{GEV} with $\epsilon=780.6261 m,\, \beta=0.7623 m,\, \gamma=-1.5476$\\ \hline
$QV$ & $C_{QV}(v_{1},v_{2})$ & Gumbel copula with $\theta = 3.1378$\\ \hline
$QVL$ & $C_{QVL}=v_{3}C_{QV}(v_{1},v_{2})$ & Nesting using independent copula \\ \hline\hline
\end{tabular}}
\end{center}
\caption{\normalsize{Model description given by \cite{setal}, changing to a \textit{GEV} distribution the modelization of $L$}}\label{table:ceppo}
\end{table}

The authors have used the quantile surfaces associated to this model to extend the notion of return period to the multivariate setting. Assuming the previous model as appropriate, we now perform a Monte Carlo simulation with a large sample size to  compare the multivariate extreme detection between the classical direction $\mathbf{e}$ (direction of the survival function) and the direction given by the first \textit{PCA}.

\begin{figure}[htbp]
\begin{center}
\includegraphics[height=6cm,width=6cm]{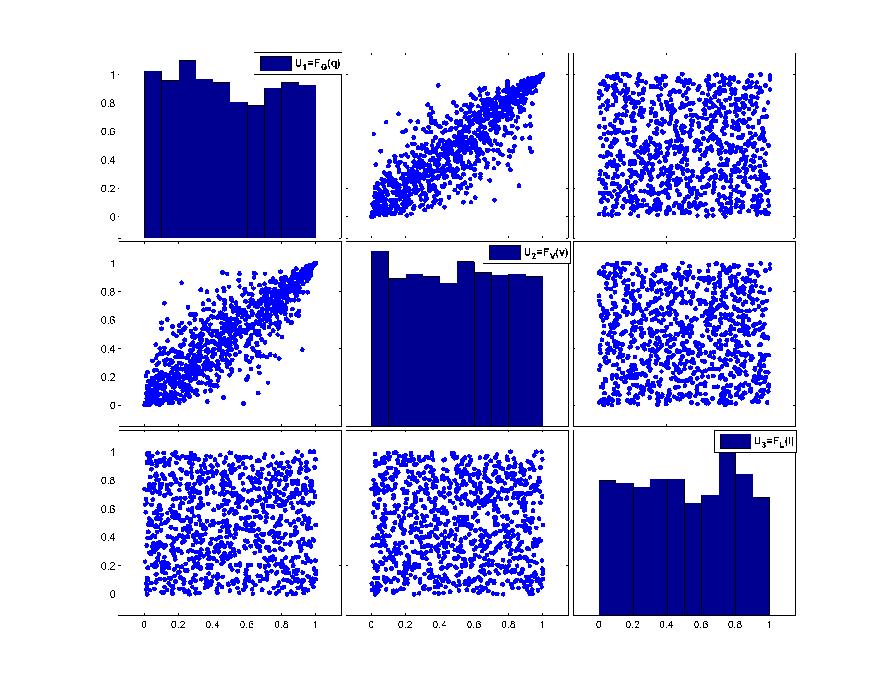}
\caption{Cross-sections of the simulated sample from the copula model}\label{fig:ceppoCop}
\vspace{-0.5cm}
\end{center}
\end{figure}

\begin{figure}[htbp]
\begin{center}
\includegraphics[height=5cm,width=5cm]{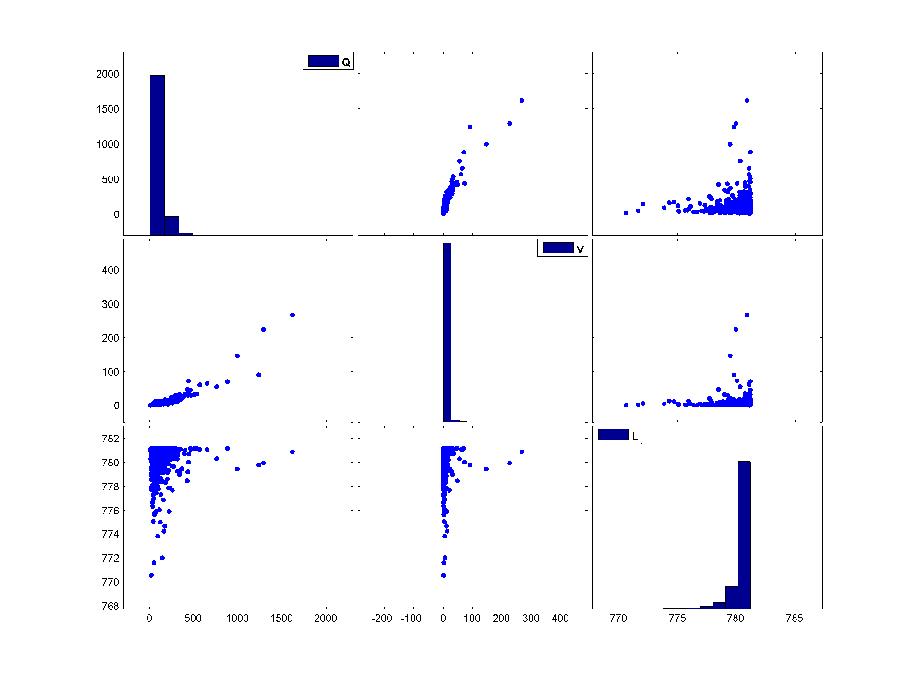}\hspace{0.5cm}\includegraphics[height=5cm,width=5cm]{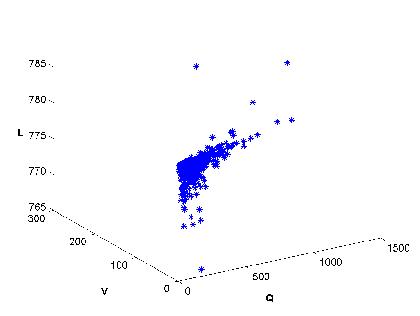}
\caption{Cross-sections and $3D-$plot of the simulated data for the model of floods}\label{fig:ceppoSim}
\vspace{-0.5cm}
\end{center}
\end{figure}

\begin{figure}[htbp]
\begin{center}
\includegraphics[height=6.5cm,width=6.5cm]{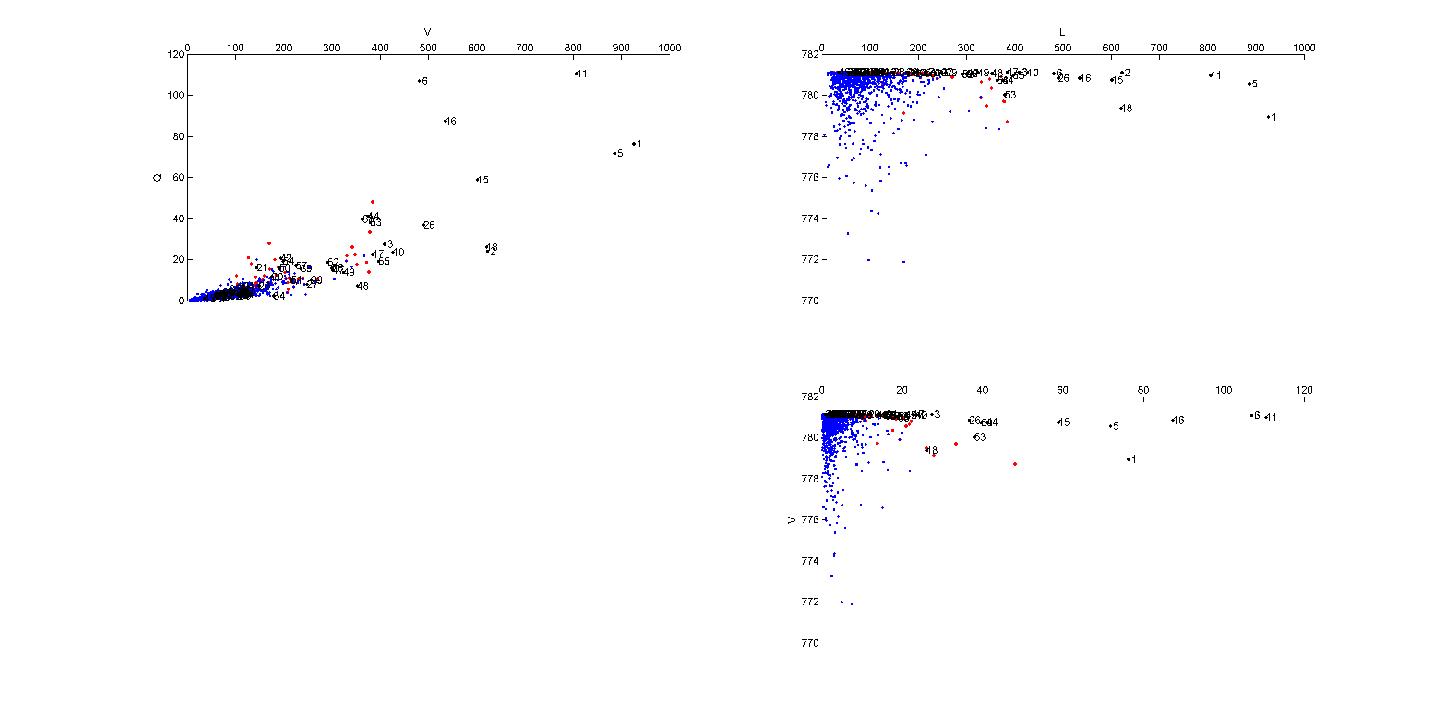}\hspace{-0.5cm}\includegraphics[height=6.5cm,width=6.5cm]{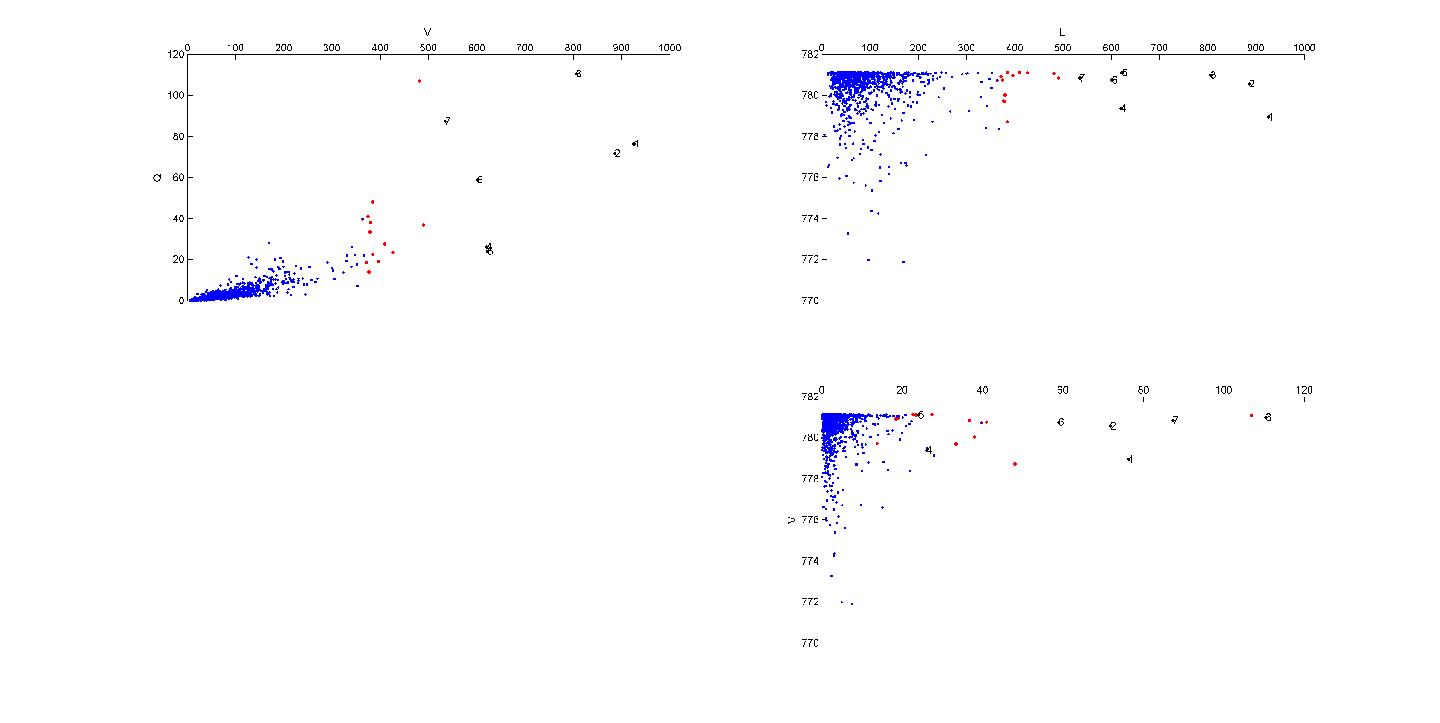}
\centerline{\hspace{1cm}(A)\;$\mathbf{u}=\mathbf{e}$\hspace{4cm} (B)\;$\mathbf{u}=$First \textit{PCA}}\ \ \\
\caption{Directional Extremes at $\alpha=1\%$}\label{fig:simCeppo_eypca}
\vspace{-0.5cm}
\end{center}
\end{figure}

Figure \ref{fig:ceppoCop} presents the cross-sections of $1000$ observations simulated from the copula model and Figure \ref{fig:ceppoSim} shows the corresponding scatterplot and $3D-$plot of the simulated data using the \textit{GEV} distributions for the marginals and Sklar's theorem to reconstruct dam behavior. Then, once the sample is generated, the extreme identification is made following the non-parametric approach at level $\alpha=1\%$ in the two directions previously mentioned. Figure \ref{fig:simCeppo_eypca} (A) illustrates the analysis considering the classical direction $\mathbf{e}$, and Figure \ref{fig:simCeppo_eypca} (B) presents the extremes obtained considering  the first \textit{PCA} direction. Both plots draw the lower or non-risky $1\%-$level sets in blue, the directional $1\%-$quantiles in red and the upper or extreme $1\%-$level sets in black.

Note that the number of extremes identified in direction $\mathbf{e}$ is significantly greater than in the first $PCA$ direction. Such a number of extremes seems excessive when a small value of $\alpha$ is considered. Also, it would not be appropriate to say that there is a \textit{$99\%$ level of confidence in the expected number of extremes in the classical direction $\mathbf{e}$}. The improvements obtained in the first $PCA$ direction are remarkable graphically.

To obtain more evidence of the advantages of the directional approach, we generate $(Q,\, V,\, L)$ triplets as inputs to operate the reservoir routing, analyzing the stress and reliability of the dam after long-time horizons of $1000-$years long. This was done similarly to \cite{demichele1}. In particular, each couple (Q, V) is transformed into a triangular flood hydrogragh of volume V and maximum peak Q, with base time $T_{b}=2V/Q$, time of rise $T_{p} = T_{b}/2.67$, and time of recession $T_{r} = 1.67 T_{p}$, (see e.g., [\cite{chow}, pg. 229]). $L$ is the water level in the dam at the beginning of the flood event. We operate the reservoir routing of flood hydrographs (see for details [\cite{bras}, pg. 475-478]) considering as outlets only the uncontrolled spillways, and checking if the spillways are capable of disposing the flood events without overtopping the dam crest.

\begin{figure}[htbp]
\begin{center}
\includegraphics[height=4cm,width=6cm]{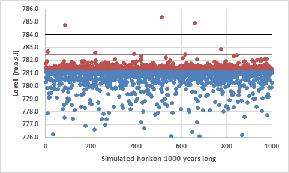}\hspace{0.5cm}\includegraphics[height=4cm,width=6cm]{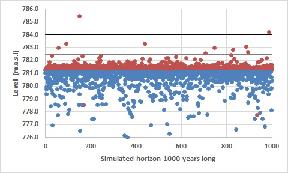}
\caption{Two (out of 100) examples of dam simulation triplets ($Q,V,L$)}\label{fig:damSim}
\vspace{-0.5cm}
\end{center}
\end{figure}

Figure \ref{fig:damSim} presents two examples of the results after the simulation of dam behavior. In the images, it is possible to see the level of the dam spillway $(781.50\, m.a.s.l.)$ which is the virtual line drawn between the maximum levels occurred (red points) and the initial levels (blue points).  Also shown are the lines defining the maximum level $(782.50\, m.a.s.l.)$ and the dam crest $(784.00\, m.a.s.l.)$. Therefore, all the points between the maximum regulation level and the dam crest are considered as risky events and those points above the dam crest are considered catastrophic events. We have done $100$ simulations of 1000-years long and the conclusion  in all of them is that the $\textit{PCA}$ directional analysis captures better the critical events, i.e., the union of the sets of points given by the risky events and the extreme or catastrophic events. Meanwhile the classical direction $\mathbf{e}$ identifies a huge number of such events, (the results considering the extension through distribution functions instead of the survival functions can be found in the supplementary material).

Table \ref{table:extResults} summarizes average indexes over the $100$ simulated samples analyzed in the two directions with $\alpha=1\%$. Specifically, the table describes: 1) The false positive ratio, which is the number of observations bad identified as critical over the total number of critical identifications. 2) The true positive ratio, which is the number of critical values correctly identified over the total number of real critical values  from the dam routing simulation. 3) The extremes detection ratio, which is the number of observations identified as critical over the total number of observations. 4) The true extremes ratio, which is the number of real critical values over the total number of observations.

\begin{table}[htbp]
\begin{center}
\begin{tabular}{|| l |c|c||}\hline\hline
 & Classical Direction & PCA Direction \\ \hline\hline
False Positives Ratio & $91.74\%$ & $52.49\%$  \\ \hline
True Positives Ratio &  $100\%$ & $100\%$ \\ \hline
Extremes Detection Ratio & $10.17\%$ & $1.77\%$  \\ \hline
True Extremes Ratio & $0.83\%$ & $0.83\%$  \\ \hline\hline
\end{tabular}
\end{center}
\caption{\normalsize{Results of the Directional Extreme Analysis}}\label{table:extResults}
\end{table}

The table shows that both directions identify correctly all the critical values with a $100\%$ of \textit{true positives ratio}, but the analysis using the first \textit{PCA} direction reduces significantly the $52.49\%$ \textit{false positives ratio} of detection, compared to $91.74\%$ in the classic direction. Also observe the small $1.77\%$ exceeding in the \textit{extreme detection ratio} given by the first \textit{PCA} direction with respect to the \textit{true extremes ratio} $0.83\%$, in comparison with the critical detection in the classic direction, which has a huge number of exceedances with a $10.17\%$ \textit{extremes detection ratio}.

\section{Case study: Sea storms}

This case study is based on a dataset of sea storms which are described by five variables. This dataset has been studied in \cite{demichele2} and was collected by a wave buoy at Alghero (Sardinia, Italy) for a period of 12 years: from July 1, 1989 to October 31, 2001. The variables considered in the study are:  wave height ($H$ in $m$), storm duration ($D$ in $hrs$), storm magnitude ($M$ in $m*hrs$), storm direction ($O$ in $deg$) and storm inter-arrival time ($I$ in $hrs$), which records the period of calm between two successive storms. It is assumed that sea storms can be considered independent and homogeneous events. A sea storm occurs when the wave height crosses upwards of $2$ meters and ends when the wave height stays below $2$ meters for at least $6$ consecutive hours. Specifically, the dataset counts a total of $415$ sea storms during the considered period.

\begin{figure}[htbp]
\begin{center}
\includegraphics[height=6.5cm,width=6.5cm]{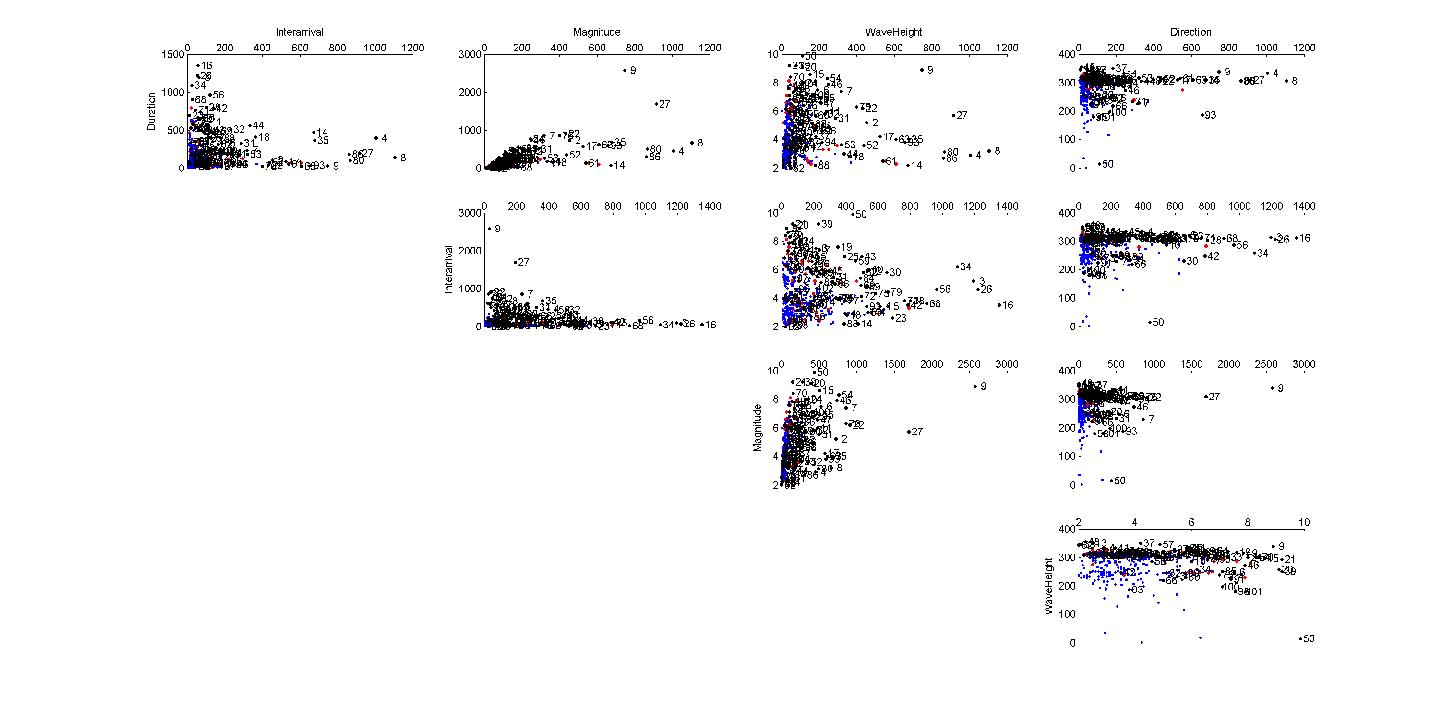}\hspace{-0.5cm}\includegraphics[height=6.5cm,width=6.5cm]{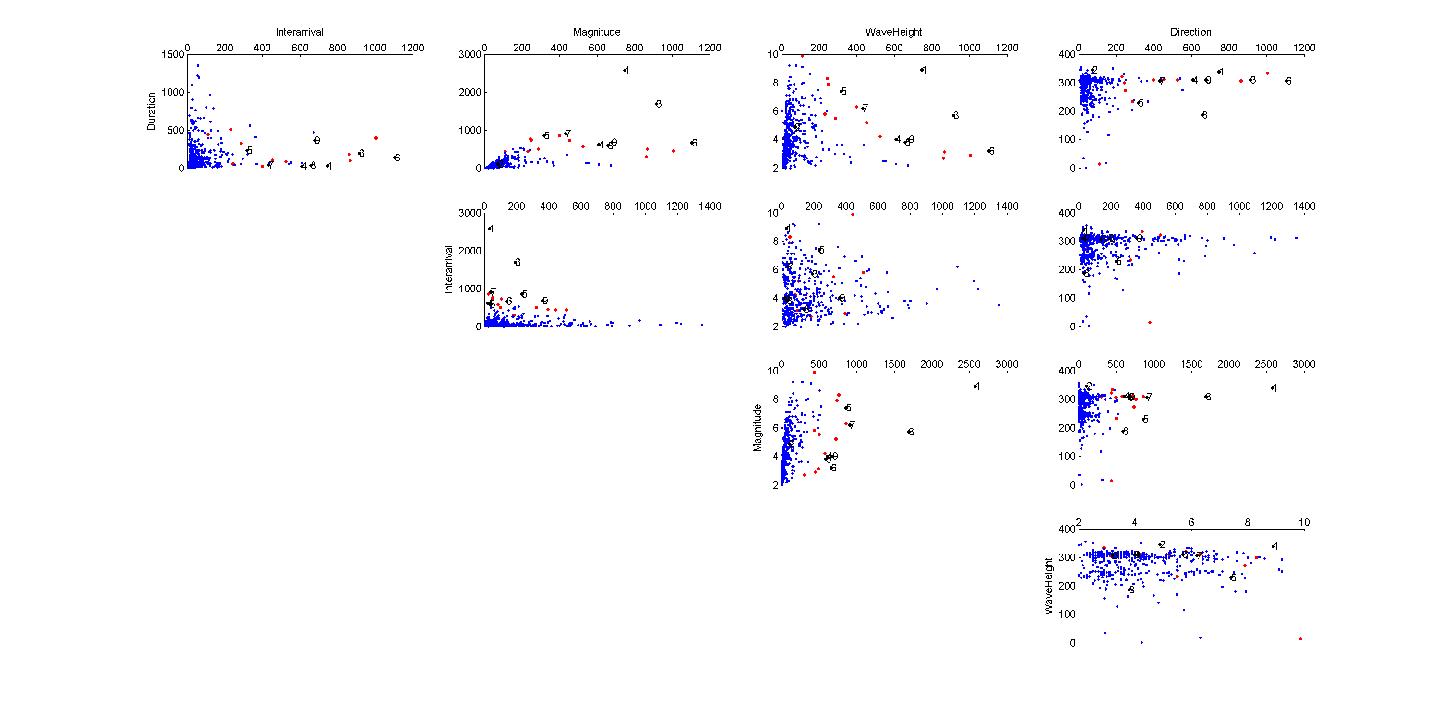}
\centerline{\hspace{1cm}(A)\;$\mathbf{u}=\mathbf{e}$\hspace{4cm} (B)\;$\mathbf{u}=$First \textit{PCA}}\ \ \\
\caption{Directional Extremes for the case study sea storms at $\alpha=1\%$}\label{fig:alg_eypca}
\vspace{-0.5cm}
\end{center}
\end{figure}

Our objective in this case study is to identify those risky events with our directional proposal in this $5-$dimensional setting, comparing the analysis in the two directions proposed in the previous case study, the classical direction $\mathbf{e}$ and the first \textit{PCA} direction. Figure \ref{fig:alg_eypca} shows the cross-sections of the sea storms dataset, where the left plot presents the identification of extremes in direction $\mathbf{e}$ for $\alpha=1\%$ (black points) and right plot shows the results associated with the first \textit{PCA} direction for the same $\alpha$ (black points). In the same way as in the previous section, the visualization of  extremes is more acceptable when the first \textit{PCA} direction is used, (the results considering the extension through distribution functions instead of the survival functions can be found in the supplementary material).

\section{Extremes based on copulas and the directional approach}

The importance of copulas have been recognized due to their capacity to capture the dependence structure of a set of random variables. Copulas are also a tool to construct families of multivariate distributions. In addition, copulas move in a compact support which guarantees  theoretical advantages. Recall for example the capability to simulate data through copulas as was shown in the case study of flood risk in a dam. Therefore, this section is devoted to introducing the directional approach to detect extremes when the dataset is modeled using copulas.

A $n-$copula $C$ is a multivariate distribution function in $[0,1]^{n}$ with  marginals equally distributed as $[0,1]-$uniforms. Also, if we consider the associated survival distribution, we get the survival copula $\bar{C}$. For example, when $n=2$, $C$ and $\bar{C}$ are linked as follows,

\begin{equation}\label{eq:survCopula}
\bar{C}(v_{1},v_{2}) = v_{1}+v_{2}-1+C(1-v_{1},1-v_{2}).
\end{equation}

The importance of modeling through copulas is due to Sklar's theorem (see Theorem \ref{teo:sklar} in the Appendix B.), since  any \textit{joint survival function} $\bar{F}$ (\textit{joint distribution function} $F$) can be obtained through its \textit{marginal survivals} $\bar{F}_{i}$, $i=1,...,n$ (\textit{marginal distributions} $F_{i}$) and the \textit{survival copula} $\bar{C}$ (\textit{copula} $C$). This representation of the models makes more feasible to obtain closed or  approximated expressions for $\mathcal{Q}_{\mathbf{X}}(\alpha,\mathbf{e})$ in \eqref{dirQ}, 
 $\mathcal{U}_{\mathbf{X}}(\alpha,\mathbf{e})$ in \eqref{upperSet} and $\mathcal{L}_{\mathbf{X}}(\alpha,\mathbf{e})$ in \eqref{lowerSet}. Thereby, in terms of survival copulas, equations \eqref{dirQ}, \eqref{upperSet} and \eqref{lowerSet} for $\mathbf{u}=\mathbf{e}$ become the following,
\begin{equation}\label{copQ}
\mathcal{Q}_{\mathbf{X}}(\alpha,\mathbf{e}) \equiv \{\mathbf{x}\in \mathbb{R}^{n} \text{ such that } x_{i}=\bar{F}_{X_{i}}^{-1}(v_{i});\ \ i=1,...,n;\ \ \mathbf{v}\in [0,1]: \  \bar{C}(\mathbf{v})=\alpha\},
\end{equation}
\begin{equation}\label{upperCop}
\mathcal{U}_{\mathbf{X}}(\alpha,\mathbf{e}) \equiv \{\mathbf{x}\in \mathbb{R}^{n} \text{ such that } x_{i}=\bar{F}_{X_{i}}^{-1}(v_{i});\ \ i=1,...,n;\ \ \mathbf{v}\in [0,1]:\  \bar{C}(\mathbf{v})<\alpha\}.
\end{equation}
\begin{equation}\label{lowerCop}
\mathcal{L}_{\mathbf{X}}(\alpha,\mathbf{e}) \equiv  \{\mathbf{x}\in \mathbb{R}^{n} \text{ such that } x_{i}=\bar{F}_{X_{i}}^{-1}(v_{i});\ \ i=1,...,n;\ \ \mathbf{v}\in [0,1]:\  \bar{C}(\mathbf{v})>\alpha\}.
\end{equation}

Most of the studies dealing with extremes detection in terms of copulas are based on definitions similar to \eqref{copQ}, \eqref{upperCop}, \eqref{lowerCop} (e.g. \cite{sydm, grimaldi2}), which are focused on the direction given by the survival function (or those associated to the parameters $(1-\alpha,-\mathbf{e})$, which are considering distribution functions). However, \cite{chebana2} use the directions $\left\{(1/\sqrt{2},\, -1/\sqrt{2}),\, (-1/\sqrt{2},\, 1/\sqrt{2})\right\}$ when negative dependent bivariate models are considered. Indeed, they consider copulas associated to a random vector, but rotated $90$ and $270$ degrees, which can be done taking advantage of the relationships between the corresponding copula $C$ and the following expressions, 
\[C_{90}(v_{1},v_{2})=u_{1}+C(v_{1},1-v_{2})\quad\text{and}\quad C_{270}(v_{1},v_{2})=v_{2}+C(1-v_{1},v_{2}).\]

These considerations highlight the need to include directions in the copula approach. Thus, the goal of this section is to include the general directional setting to the copula approach and to describe a directional extreme detection method based on copulas, although we will also show the drawbacks of the procedure with some illustrative simulations. The following result shows how the directional approach can be implemented using copulas. 
\begin{Proposition}\label{prop:pcaRel}
\textit{Let $\mathbf{u}$ be fixed, then the directional quantiles and the associated upper and lower level sets of a random vector $\mathbf{X}$ (defined in \eqref{dirQ}, \eqref{upperSet} and \eqref{lowerSet}) are the same as those obtained by applying the copula method (summarized in \eqref{copQ}, \eqref{upperCop} and \eqref{lowerCop}) to the random vector $R_{\mathbf{u}}\mathbf{X}$, where $R_{\mathbf{u}}$ is the rotation matrix in (\ref{eq:convexcone}).}
\end{Proposition}

The proof is given in Appendix B. As a conclusion, the directional analysis can be done theoretically using copula models but over the pre-rotated random vector. Some examples to illustrate Proposition \ref{prop:pcaRel} on the bivariate field are provided below. Indeed, $n=2$ can be considered as the foundation of the nesting copula procedures used in the literature to confront the problem of large dimensions: \textit{nested copula method} (see \cite{demichele2, grimaldi1}) and  \textit{Pair-copula construction}, also called the \textit{Vine copula method} (see \cite{grimaldi2}). 

Let $\mathbf{X}=(X_{1},X_{2})$ be a bivariate vector with Gaussian survival marginals $\bar{F}_{1}$, $\bar{F}_{2}$ with parameters $\mu_{1}$, $\sigma_{1}^{2}$ and $\mu_{2}$, $\sigma_{2}^{2}$ respectively. We also assume that  $\mathbf{X}$  satisfies a Gaussian survival copula $\bar{C}$ with Pearson's correlation coefficient $\rho$. Note that, 

\begin{equation}\label{eq:meanVar}
\begin{aligned}
\mu_{i} &= \int_{-\infty}^{\infty}x_{i}dF_{i}= \int_{0}^{1}F_{i}^{-1}(u_{i})du_{i}\\
\sigma_{i}^{2} &= \int_{-\infty}^{\infty}(x_{i}-\mu_{i})^{2}dF_{i}= \int_{0}^{1}(F_{i}^{-1}(u_{i})-\mu_{i})^{2}du_{i},
\end{aligned}
\end{equation}

where $F=1-\bar{F}$, for all $i = 1,2$ and denote the covariance matrix of $\mathbf{X}$ by,

\[\Sigma=\begin{pmatrix}
	\sigma_{1}^{2} & \sigma_{1}\sigma_{2}\rho\\
	\sigma_{1}\sigma_{2}\rho & \sigma_{2}^{2}
	\end{pmatrix}.\]

It is well known that the Gaussian  copula  is closed under orthogonal transformations. Then, for any direction $\mathbf{u}$, $R_{\mathbf{u}}\mathbf{X}$ also  holds a Gaussian survival copula $\bar{C}^{\mathbf{u}}$ with Pearson's correlation coefficient given by
\begin{equation}\label{eq:rhoRot}
	\rho^{\mathbf{u}}=\left[R_{\mathbf{u}}\Sigma R_{\mathbf{u}}'\right]_{12},
\end{equation}

and Gaussian survival marginals $\bar{F}^{\mathbf{u}}_{1}$, $\bar{F}^{\mathbf{u}}_{2}$ with parameters
\begin{equation}\label{eq:margRot}
	\mu_{1}^{\mathbf{u}}=\left[R_{\mathbf{u}}\begin{pmatrix}
	\mu_{1}\\
	\mu_{2}
	\end{pmatrix}\right]_{11}, \sigma_{1}^{2\,\mathbf{u}}=\left[R_{\mathbf{u}}\Sigma R_{\mathbf{u}}'\right]_{11}\quad \text{and}\quad
	\mu_{2}^{\mathbf{u}}=\left[R_{\mathbf{u}}\begin{pmatrix}
	\mu_{1}\\
	\mu_{2}
	\end{pmatrix}\right]_{21}, \sigma_{2}^{2\, \mathbf{u}}=\left[R_{\mathbf{u}}\Sigma R_{\mathbf{u}}'\right]_{22},
\end{equation}

where $[\cdot]_{ij}$ is the $i$, $j$ position in a matrix.

\begin{figure}[htbp]
\begin{center}
\includegraphics[height=30mm,width=30mm]{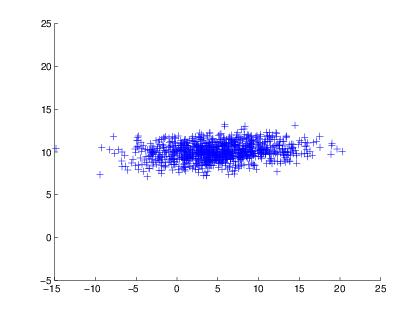}\hspace{1cm}
\includegraphics[height=30mm,width=30mm]{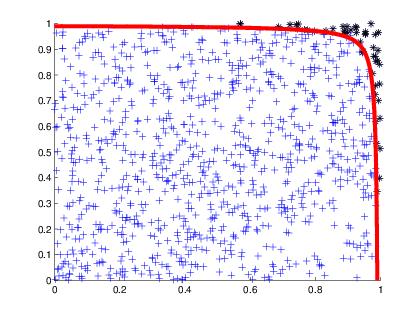}\hspace{1cm}\includegraphics[height=30mm,width=30mm]{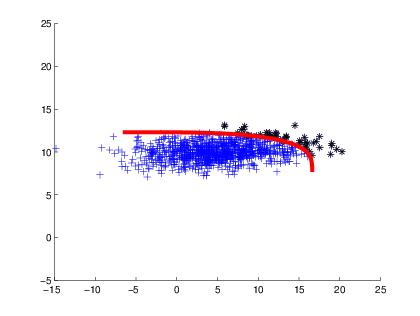}\ \ \\
\centerline{(A)\;Original data\hspace{1.5cm} (B)\;Original data in \hspace{1.5cm} (C)\;Original data}\ \ \\
\vspace{-0.5cm}
\centerline{the copula space}\ \ \\
\includegraphics[height=30mm,width=30mm]{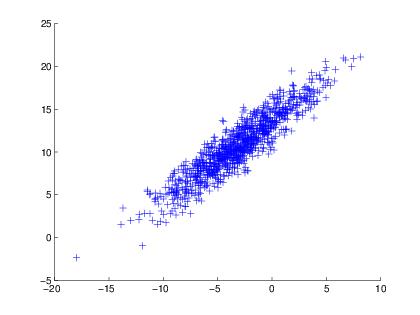}\hspace{1cm}
\includegraphics[height=30mm,width=30mm]{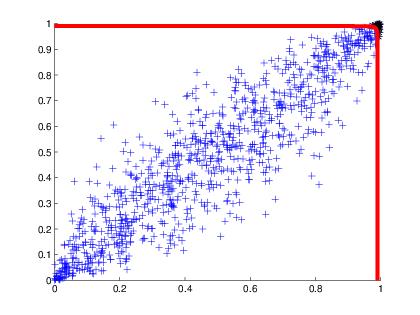}\hspace{1cm}\includegraphics[height=30mm,width=30mm]{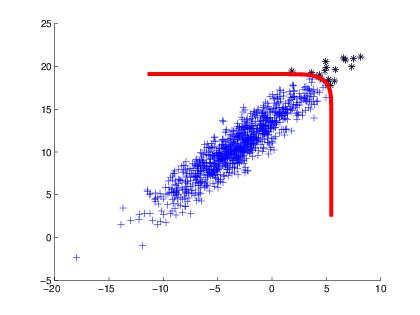}\ \ \\
\centerline{(D)\;Rotated data\hspace{1.2cm} (E)\;Rotated data in\hspace{1.2cm} (F)\;Rotated data}\ \ \\
\vspace{-0.5cm}
\centerline{the copula space}\ \ \\
\caption{Top: theoretical results in direction $\mathbf{e}$; Bottom: theoretical results in direction $\mathbf{e}$ for the rotation of the data given by the first \textit{PCA} direction}\label{fig:normCase}
\vspace{-0.5cm}
\end{center}
\end{figure}

Now, we fix the parameters $\mu_{1} = 5$, $\sigma_{1}^{2}=25$, $\mu_{2} = 10$, $\sigma_{2}^{2}=1$ and $\rho = 0.2$ to illustrate the extreme detection  through copulas.  Figure \ref{fig:normCase} summarizes the results. The three top plots describe the procedure  in the classical direction $\mathbf{e}$ for $\alpha=1\%$ and the three bottom plots describe the results in the first \textit{PCA} direction for the same $\alpha$. Figure \ref{fig:normCase}(A) shows the simulated data from the Gaussian model previously described, Figure \ref{fig:normCase}(B) plots the copula space of the data (Gaussian copula) and the theoretical  $\alpha-$quantile (red), the lower (blue) and upper (black) level sets following the equations \eqref{copQ}, \eqref{upperCop} and \eqref{lowerCop}. Finally, Figure \ref{fig:normCase}(C) shows the corresponding results once the original space of the data is recovered through the inverse of the marginal survivals (all the colors have the same meaning as in Figure \ref{fig:normCase}(B)).

\begin{figure}[htbp]
\begin{center}
\includegraphics[height=40mm,width=40mm]{normData_e_1.jpg}\hspace{1cm}
\includegraphics[height=40mm,width=40mm]{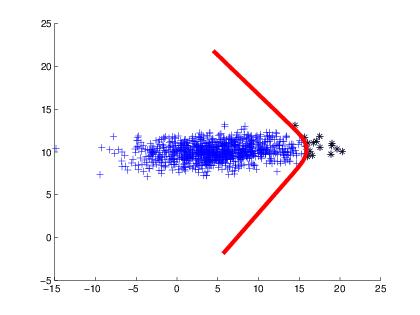}\ \ \\
\centerline{\hspace{0.2cm}(A)\; Classical direction $\mathbf{e}$\hspace{1.3cm} (B)\; First \textit{PCA} direction}\ \ \\
\caption{Comparison of the identification of extremes in directions $\mathbf{e}$ and first \textit{PCA} (black points)}\label{fig:normResults}
\vspace{-0.5cm}
\end{center}
\end{figure}

In a similar way, but for the first \textit{PCA} direction, Figure \ref{fig:normCase}(D)  shows the pre-rotated data, Figure \ref{fig:normCase}(E)  plots  the copula space of the rotated data and the extreme detection in  the copula space, and Figure \ref{fig:normCase}(F) displays the extremes in the rotated space after applying the inverse of the rotated survival marginals. In order to compare the results in both directions, Figure \ref{fig:normResults}(A) shows the extremes considering the  direction $\mathbf{e}$ and Figure \ref{fig:normResults}(B) shows extremes in the first \textit{PCA} direction undoing the rotation $R_{\mathbf{u}}$. Graphically, the differences in the two directions are obvious and the extremes detected using the first \textit{PCA} direction look more realistic since they are more congruent with the shape of the data. 

The Gaussian copula is a toy example where the directional approach can be theoretically extended to the classical copula procedure. However, the usual fact is that the knowledge of the copula and the marginals associated to a random vector $\mathbf{X}$ does not imply knowing the copula and the marginals over a rotation of the random vector. Therefore, a disadvantage of the directional copula approach is that it increases the computational cost when one decides to consider another direction of analysis different from $\mathbf{e}$. 

\begin{figure}[htbp]
\begin{center}
\includegraphics[height=30mm,width=30mm]{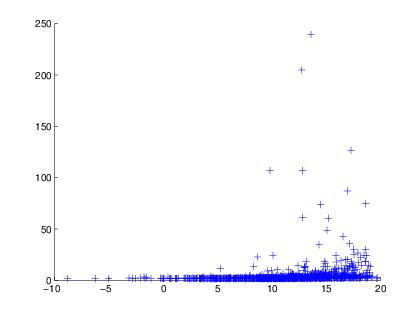}\hspace{1cm}
\includegraphics[height=30mm,width=30mm]{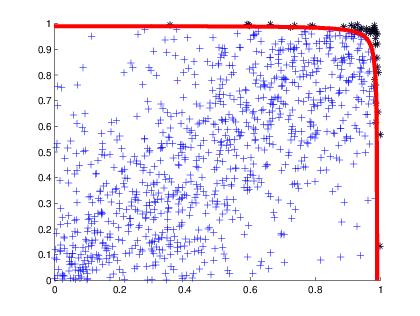}\hspace{1cm}\includegraphics[height=30mm,width=30mm]{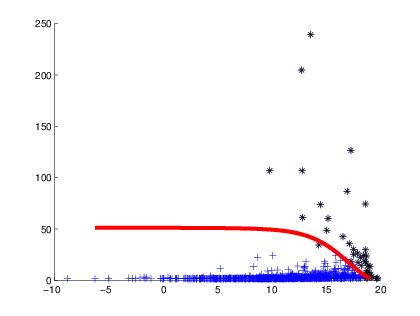}\ \ \\
\centerline{(A)\;Original data\hspace{1.5cm} (B)\;Original data in \hspace{1.5cm} (C)\;Original data}\ \ \\
\vspace{-0.5cm}
\centerline{the copula space}\ \ \\
\includegraphics[height=30mm,width=30mm]{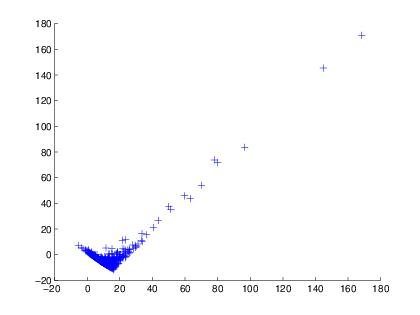}\hspace{1cm}
\hspace{4cm}
\includegraphics[height=30mm,width=30mm]{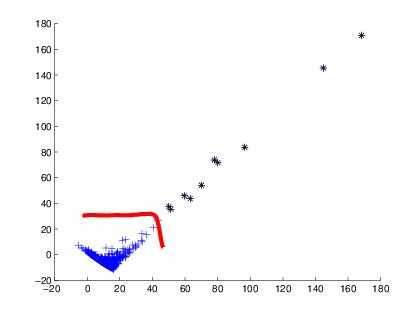}
\centerline{(D)\;Rotated data\hspace{1.2cm} (E)\;Rotated data in\hspace{1.2cm} (F)\;Rotated data}\ \ \\
\centerline{the copula space}\ \ \\
\caption{Top: theoretical results in direction $\mathbf{e}$; Bottom: non-parametric approach in direction $\mathbf{e}$ for the rotation of the data given by the first \textit{PCA} direction}\label{fig:depPos}
\vspace{-0.5cm}
\end{center}
\end{figure}

For example, let us consider a \textit{Frank} copula, and marginal distributions belonging to the \textit{GEV} family (see Appendix B. for more details about the formulations of a Frank copula and GEV distributions). Firstly, we have assumed positive dependence in the model by setting a Frank survival copula with dependence parameter $\theta=5$ and \textit{GEV} marginals with parameters $\beta_{1}=5$, $\epsilon_{1}=10$, $\gamma_{1}=-1/2$, $\beta_{2}=1/2$, $\epsilon_{2}=2$ and $\gamma_{2}=1$.

\begin{figure}[htbp]
\begin{center}
\includegraphics[height=40mm,width=40mm]{depPosData_e_1.jpg}\hspace{1cm}
\includegraphics[height=40mm,width=40mm]{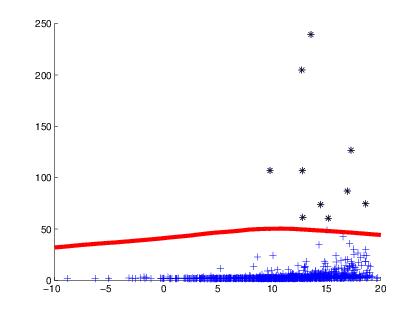}\ \ \\
\centerline{\hspace{0.2cm}(A)\; Classical direction $\mathbf{e}$\hspace{1.3cm} (B)\; First \textit{PCA} direction}\ \ \\
\caption{Comparison of the identification of extremes in directions $\mathbf{e}$ and first \textit{PCA} (black points)}\label{fig:posResults}
\vspace{-0.5cm}
\end{center}
\end{figure}

Figure \ref{fig:depPos}(A, B, C) show the classical theoretical procedure used with copulas for $\alpha=1\%$ and direction $\mathbf{e}$ with the same meaning as in Figure \ref{fig:normCase}(A,B,C). However, Figure \ref{fig:depPos}(D, E, F) plot  the analysis for the same $\alpha$, but using the non-parametric approach in direction $\mathbf{e}$ over the pre-rotated data under the rotation $R_{\mathbf{u}}$ given by the first \textit{PCA} direction. Figure \ref{fig:depPos}(D) shows the data in the rotated space, Figure \ref{fig:depPos}(E) is empty due to the absence of theoretical evidence of the copula after the rotation of the data. Note that a possibility to fill the empty figure is to apply the non-parametric directional procedure presented in Section \ref{subsec:dirApp}, but to the non-parametric copula of the rotated data (see \cite{caperaa}), since Proposition \ref{prop:pcaRel} guarantees the theoretical equivalence. However, the directional approach has the advantage that the extremes can be obtained without considering the copula space of the rotated data as is shown in the identification presented in Figure \ref{fig:depPos}(F).

\begin{figure}[htbp]
\begin{center}
\includegraphics[height=30mm,width=30mm]{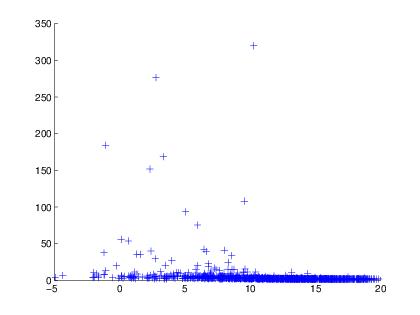}\hspace{1cm}
\includegraphics[height=30mm,width=30mm]{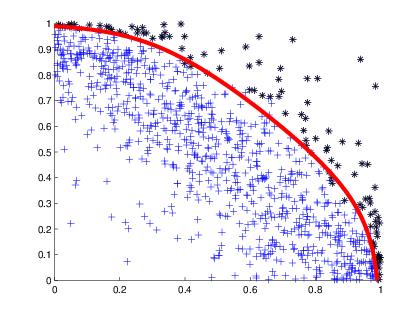}\hspace{1cm}\includegraphics[height=30mm,width=30mm]{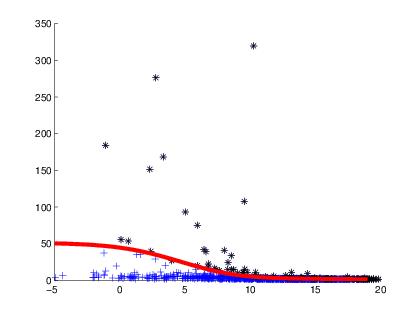}\ \ \\
\centerline{(A)\;Original data\hspace{1.5cm} (B)\;Original data in \hspace{1.5cm} (C)\;Original data}\ \ \\
\vspace{-0.5cm}
\centerline{the copula space}\ \ \\
\includegraphics[height=30mm,width=30mm]{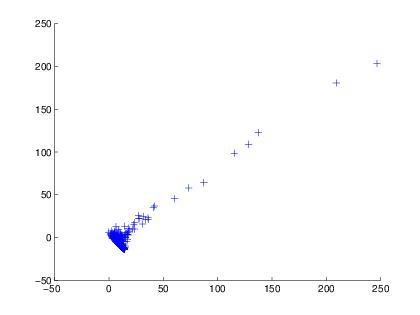}\hspace{1cm}
\hspace{4cm}
\includegraphics[height=30mm,width=30mm]{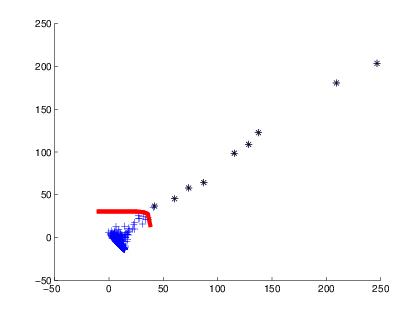}
\centerline{(D)\;Rotated data\hspace{1.2cm} (E)\;Rotated data in\hspace{1.2cm} (F)\;Rotated data}\ \ \\
\centerline{the copula space}\ \ \\
\caption{Top: theoretical results in direction $\mathbf{e}$; Bottom: non-parametric approach in direction $\mathbf{e}$ for the rotation of the data given by the first \textit{PCA} direction}\label{fig:depNeg}
\vspace{-0.5cm}
\end{center}
\end{figure}

To compare the detected extremes, Figure \ref{fig:posResults} displays in black those points considered as extremes in both directions, once the rotation of the data is undone in the case of the first \textit{PCA} direction. The large number of points identified as extremes in the case of the classical direction $\mathbf{e}$ with $\alpha=1\%$ can be observed, when many of these identified observations could be considered as regular observations. Meanwhile using the first \textit{PCA} direction, the number of extremes is considerably reduced and they appear more reasonable. 

\begin{figure}[htbp]
\begin{center}
\includegraphics[height=40mm,width=40mm]{depNegData_e_1.jpg}\hspace{1cm}
\includegraphics[height=40mm,width=40mm]{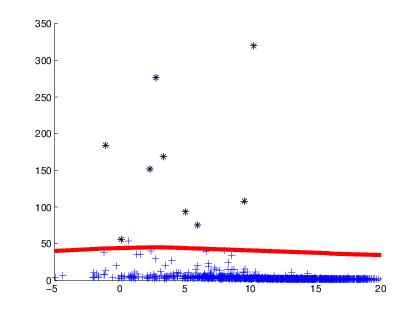}\ \ \\
\centerline{\hspace{0.2cm}(A)\; Classical direction $\mathbf{e}$\hspace{1.3cm} (B)\; First \textit{PCA} direction}\ \ \\
\caption{Comparison of the identification of extremes in directions $\mathbf{e}$ and first \textit{PCA} (black points)}\label{fig:negResults}
\vspace{-0.5cm}
\end{center}
\end{figure}

To conclude this section, we consider a model with negative dependence. In this case, $\alpha$ is again $1\%$, but the parameter of dependence in the Frank survival copula is $\theta=-8$ and we use the same \textit{GEV} marginals as in the previous example. Figure \ref{fig:depNeg} shows the outputs in the same framework as Figure \ref{fig:depPos}, and Figure \ref{fig:negResults} shows the contrast between the classical and the first \textit{PCA} directions for the detection of extremes. Once again, we can observe a better pattern of extreme recognition by considering the alternative direction of analysis that we propose, the first \textit{PCA} direction, (in the supplementary material are analyzed all the examples described in this Section but considering the extension through distribution functions).

\section{Conclusions}

In this paper we propose a directional multivariate extreme identification procedure based on the notion of directional multivariate quantile. A non-parametric implementation feasible in high dimensions is also presented. We have proposed a directional inclusion to the classical extreme detection procedure based on copulas and we have analyzed simulated and real scenarios where the advantages of using different directions to detect extremes is evident. Specifically, \textit{Principal Component Analysis} has been tested as a method to choose a suitable direction of analysis that offers a reasonable number of points identified as extremes, but more importantly, the locations of those identifications are more in the "{\it atypical zone}", if one looks at the cloud of observations and its shape.  Finally, we have highlighted the advantages and disadvantages of the directional non-parametric approach and the directional copula procedure.

\section*{Acknowledgments}

This research was partially supported by a Spanish Ministry of Economy and Competitiveness grant ECO2012-38442.

\section*{Appendix A.}

\begin{Definition} (Definition 2.2 in \cite{torres1st}) \label{def:orthant}
An oriented orthant in $\mathbb{R}^{n}$ with vertex $\mathbf{x}$ in the direction $\mathbf{u}$ is defined by,
\begin{equation}
	\mathfrak{C}_{\mathbf{x}}^{R_{\mathbf{u}}}=\{\mathbf{z}\in \mathbb{R}^{n}:R_{\mathbf{u}}(\mathbf{z}-\mathbf{x})\geq 0\}.
\end{equation}
where $\mathbf{u}\in\{\mathbf{v}\in\mathbb{R}^{n}:||\mathbf{v}||=1\}$ and $R_{\mathbf{u}}$ is an orthogonal matrix such that $R_{\mathbf{u}}\mathbf{u}=\mathbf{e}$, with $\mathbf{e}=\frac{\sqrt{n}}{n}[1,...,1]'$.\\
\end{Definition}
Let $\mathbf{u}$ be a unit vector with non-null components and let $M_{\mathbf{u}}$ and $M_{\mathbf{e}}$ be matrices defined as,
\begin{equation}\label{eq:basicMat}
M_{\mathbf{u}} = [\mathbf{u},\ \ sgn(u_{2})\mathbf{e}_{2},\ \ \cdots,\ \ sgn(u_{n})\mathbf{e}_{n}]\quad\text{ and }\quad M_{\mathbf{e}} = [\mathbf{e},\ \ \mathbf{e}_{2},\ \ \cdots,\ \ \mathbf{e}_{n}],
\end{equation}
where $u_{i}$, $i=1,...,n$ is the $i-$th component of $\mathbf{u}$, $sgn(\cdot)$ is the scalar sign function and $\mathbf{e}_{i}$ is the vector with all its components equal to zero except the $i-$th component equal to one. Note that  the hypothesis of $u_{i}\neq 0$, $i=1,...,n$ guarantees that  $M_{\mathbf{u}}$ always is a matrix  of rank $n$. Now, we consider the QR decomposition of  $M_{\mathbf{u}}$ and $M_{\mathbf{e}}$ (see e.g. \cite{horn}, Ch. 2),

\[M_{\mathbf{u}} = Q_{\mathbf{u}}T_{\mathbf{u}}\quad\text{ and }\quad M_{\mathbf{e}} = Q_{\mathbf{e}}T_{\mathbf{e}},\]

such that $T_{\mathbf{u}}$ and $T_{\mathbf{e}}$ are triangular matrices with positive diagonal elements, and $Q_{\mathbf{u}}$ and $Q_{\mathbf{e}}$ are orthogonal matrices. Note that these decompositions are unique due to both the full rank of $M_{\mathbf{u}}$ and $M_{\mathbf{e}}$, and the hypothesis of positive diagonal elements in  $T_{\mathbf{u}}$ and $T_{\mathbf{e}}$ (see e.g. \cite{horn}, Theorem 2.1.14, p.g. 89).

Also, the first columns on $Q_{\mathbf{u}}$ and $Q_{\mathbf{e}}$ are the same as in $M_{\mathbf{u}}$ and $M_{\mathbf{e}}$; that is,  $\mathbf{u}$ and $\mathbf{e}$ respectively. Therefore,
$Q_{\mathbf{e}}\mathbf{e}_{1}=\mathbf{e}$ and $Q_{\mathbf{u}}\mathbf{e}_{1}=\mathbf{u}$ and thus, $\left(Q_{\mathbf{e}}Q'_{\mathbf{u}}\right) \mathbf{u}= \mathbf{e}$. This decomposition motivates the definition of the $QR$ oriented orthant where the rotation matrix is unique. \\

\begin{Definition}\label{def:orthantQR}
The QR oriented orthant with vertex $\mathbf{x}$ in direction $\mathbf{u}$ is an oriented orthant satisfying  $R_{\mathbf{u}}=Q_{\mathbf{e}}Q'_{\mathbf{u}}$.
\end{Definition}

\section*{Appendix B.}

\begin{Theorem}[\textbf{Sklar's theorem}]\label{teo:sklar}
\textbf{Sklar's theorem:} Let $F$ be a $n$-dimensional distribution function with marginals $F_{1},\cdots,F_{n}$. Then, there exists a $n$-copula $C$ such that for all $\mathbf{x}\in \bar{\mathbb{R}}^{n}$,

\begin{equation}\label{eq:relCopula}
	F(x_{1},\cdots,x_{n})=C(F_{1}(x_{1}),\cdots,F_{n}(x_{n})).
\end{equation}

If $F_{1},...,F_{n}$ are  continuous distribution functions, then $C$ is unique. On the other hand, $C$ is only defined on $Ran(F_{1})\times \cdots\times Ran(F_{n})$. Conversely, if $C$ is a $n$-copula and $F_{1},...,F_{n}$ are distributions functions, then the function $F$ defined by (\ref{eq:relCopula}) is a $n$-dimensional distribution function with marginals $F_{1},...,F_{n}$.\\
\end{Theorem}

Also, it is possible to express this result linking the joint survival function of the random vector through a survival copula, with the survival marginals by the equation,

\begin{equation}\label{eq:relSurCopula}
	\bar{F}(x_{1},\cdots,x_{n})=\bar{C}(\bar{F}_{1}(x_{1}),\cdots,\bar{F}_{n}(x_{n})).
\end{equation}

In the literature there are many classes or families of copulas and they have become a powerful tool for modeling  practical situations in the multivariate framework where there is relevant joint information. For a deeper discussion of  copula theory, we refer the reader to \cite{nelsen} and \cite{nature}. The formulation of the two families of bivariate copulas used in the paper is the following:\\

\textbf{Gaussian Copula:}
The Gaussian copula is given by the expression
\begin{equation}\label{eq:normCop}
	C(v_{1},v_{2}) = \Phi_{\rho}(\Phi_{1}^{-1}(v_{1}),\Phi_{2}^{-1}(v_{2})),
\end{equation}
where $\Phi_{\rho}$ is a bivariate standard Gaussian distribution with Pearson's correlation coefficient $\rho$, $\Phi_{1}^{-1}$ and $\Phi_{2}^{-1}$ the pseudo-inverse of Gaussian univariate distributions with parameters $\mu_{1}$, $\sigma_{1}^{2}$ and $\mu_{2}$, $\sigma_{2}^{2}$ respectively.\\

\textbf{The Frank Copula:}
An Archimedean copula with the following bivariate expression:

\begin{equation}\label{eq:frankCop}
	 C_{\theta}(v_{1},v_{2})=-\frac{1}{\theta}ln\left(1+\frac{(e^{-\theta v_{1}}-1)(e^{-\theta v_{2}}-1)}{e^{-\theta}-1}\right),
\end{equation}
where $\theta\in\mathbb{R}/\{0\}$.\\

Finally, we summarize a set of univariate distributions quite useful in extreme value analysis (see \cite{kotz}), which is called \textit{Generalized Extreme Value Distributions}, \textit{GEV}. Then a distribution belongs to the \textit{GEV} family if it follows the structure:\\

\textbf{GEV Distributions:}
\begin{equation}\label{eq:gev}
	F_{X}(x)=\begin{cases}
	exp\left\{-\left[1+\gamma\left(\frac{x-\epsilon}{\beta}\right)\right]^{-\frac{1}{\gamma}}\right\} \qquad \text{if } \gamma \neq 0,\\
	exp\left\{-exp\left\{-\left(\frac{x-\epsilon}{\beta}\right)\right\}\right\} \qquad \text{if } \gamma = 0,
	\end{cases}
\end{equation}

where $\beta>0$, $\epsilon,\ \ \gamma\in\mathbb{R}^{n}$ are the scale, location and shape parameters respectively. For a good introduction to univariate extreme analysis as well as the multivariate approach using copulas and the applications, we refer to \cite{nature}.

\begin{Prop}
\textit{Let $\mathbf{u}$ be fixed, then the directional quantiles and the associated   upper and lower level sets of a random vector $\mathbf{X}$ (defined in \eqref{dirQ}, \eqref{upperSet} and \eqref{lowerSet}) are the same as those obtained by applying the copula method (summarized in \eqref{copQ}, \eqref{upperCop} and \eqref{lowerCop}) to the random vector $R_{\mathbf{u}}\mathbf{X}$, where $R_{\mathbf{u}}$ is the rotation matrix in (\ref{eq:convexcone}).}
\end{Prop}

\begin{Proof}
First note  that any  analysis using the information of the survival or the distribution functions for a random vector $\mathbf{X}$ through copulas is equivalent to the analysis in the set of classic directions $\{\mathbf{e},-\mathbf{e}\}$, i.e., the copula quantile analysis is always done in those directions. Moreover, once $\mathbf{u}$ is fixed, Sklar's theorem provides  the following relationships between the random vector $R_{\mathbf{u}}\mathbf{X}$ and the copulas $C_{\mathbf{u}}$, $\bar{C}_{\mathbf{u}}$ for any pre-rotation $R_{\mathbf{u}}$,

\begin{align}\label{eq:dirSklar1}
\mathbb{P}\left[\mathbf{X}\in\mathfrak{C}_{\mathbf{x}}^{\mathbf{u}}\right]&=\bar{F}_{R_{\mathbf{u}}\mathbf{X}}(\mathbf{x})=\bar{C}_{\mathbf{u}}\left(\bar{F}_{[R_{\mathbf{u}}\mathbf{X}]_{1}}(x_{1}),\cdots, \bar{F}_{[R_{\mathbf{u}}\mathbf{X}]_{n}}(x_{n})\right),\\ \label{eq:dirSklar2}
\mathbb{P}\left[\mathbf{X}\in\mathfrak{C}_{\mathbf{x}}^{-\mathbf{u}}\right]&=F_{R_{-\mathbf{u}}\mathbf{X}}(\mathbf{x})=C_{\mathbf{u}}\left(F_{[R_{-\mathbf{u}}\mathbf{X}]_{1}}(x_{1}),\cdots, F_{[R_{-\mathbf{u}}\mathbf{X}]_{n}}(x_{n})\right),
\end{align}

where $\bar{F}_{[R_{\mathbf{u}}\mathbf{X}]_{i}}(x_{i})$, $F_{[R_{-\mathbf{u}}\mathbf{X}]_{i}}(x_{i})$, $i=1,...,n$ are respectively the marginal survival and distribution functions of the rotated random vector $R_{\mathbf{u}}\mathbf{X}$. Hence we get the defined sets in (\ref{dirQ}), (\ref{upperSet}) and (\ref{lowerSet}), by applying the inverse of the rotation $R_{\mathbf{u}}$ over the results of the equations (\ref{copQ}), (\ref{upperCop}) and (\ref{lowerCop}) through the relationship in \ref{eq:dirSklar1}.

\end{Proof}

\section*{Supplementary Material}

\subsection{Methodology}

The purpose of this supplementary material is to show the differences of detecting multivariate extremes when the  \textit{distribution function} is used instead of the 
 \textit{survival function}. Recall that in the notion QR \textit{oriented orthant} given in Section 2.1 of the paper,  the value of the distribution function $F$ of a random vector $\mathbf{X}$  evaluated at some point $\mathbf{x}\in \mathbf{R}^{n}$ is the same as the probability of the oriented orthant in direction $-\mathbf{e}$ and vertex $\mathbf{x}$. On  the other hand, the value of its survival function $\bar{F}$ at $\mathbf{x}$ agrees with the probability of the oriented orthant in direction $\mathbf{e}$ and vertex $\mathbf{x}$. In the directional framework,   the same discussion holds  for $R_{\mathbf{u}}\mathbf{X}$. 

We also highlight that upper and lower sets are strictly related with directions due to geometrical aspects.  For instance, in the univariate setting when we are interested in minima of the variable, we focus on the left tail of the density function and 
the interpretation of an upper set is related to values less than a chosen quantile. However,  when the interest is in maxima, one look to the right tail of the density function and the upper set corresponds to values greater than the quantile, i.e., the upper set depends on the chosen direction $-1$ (\textit{distributions}) or $+1$ (\textit{survivals}). The same happens in the multivariate setting, but the complexity increases because there are infinite directions to analyze extremes and it is possible to get both approaches for upper sets for each chosen direction, due to the duality on the extension.

In the paper, we have carried out all the analysis using  survival function (see equations (2.2), (2.3) and (2.4)), but we can also consider  distribution function  easily if we consider  the pair $(1-\alpha,-\mathbf{u})$, and  the inequalities also changed in the following way,

\begin{equation}\label{dirQApp}
	\underline{\mathcal{Q}}_{\mathbf{X}}(\alpha,\mathbf{u})=\partial\{\mathbf{z}\in \mathbb{R}^{n}:\mathbb{P}\left[\mathfrak{C}_{\mathbf{z}}
	^{-\mathbf{u}}\}\right]\geq 1-\alpha\}
\end{equation}

\begin{equation}\label{upperSetApp}
		\underline{\mathcal{U}}_{\mathbf{X}}(\alpha,\mathbf{u})=\{\mathbf{z}\in \mathbb{R}^{n}:\mathbb{P}\left[\mathfrak{C}_{\mathbf{z}}
		^{-\mathbf{u}}\}\right]> 1-\alpha\}
\end{equation}

\begin{equation}\label{lowerSetApp}
		\underline{\mathcal{L}}_{\mathbf{X}}(\alpha,\mathbf{u})=\{\mathbf{z}\in \mathbb{R}^{n}:\mathbb{P}\left[\mathfrak{C}_{\mathbf{z}}
		^{-\mathbf{u}}\}\right]< 1-\alpha\}.
\end{equation}

Equation (2.5) in the paper establishes an important relationship between upper sets in both approaches, which induces that an approach through distribution functions is less conservative, because the number of points identified as extremes in each chosen direction is reduced.

In this supplementary material, we show the results under the  distribution function approach in each of the scenarios proposed in the paper, real and simulated ones, in order to present the alternative and compare the approaches. Hereafter, we refer to the analysis using equations \eqref{dirQApp}, \eqref{upperSetApp} and \eqref{lowerSetApp} evaluated at $\alpha=1\%$ as the analysis through distribution functions at level $99\%$.

\subsection{Case study: flood risk at a dam}

\begin{figure}[htbp]
\begin{center}
\includegraphics[height=6.5cm,width=6.5cm]{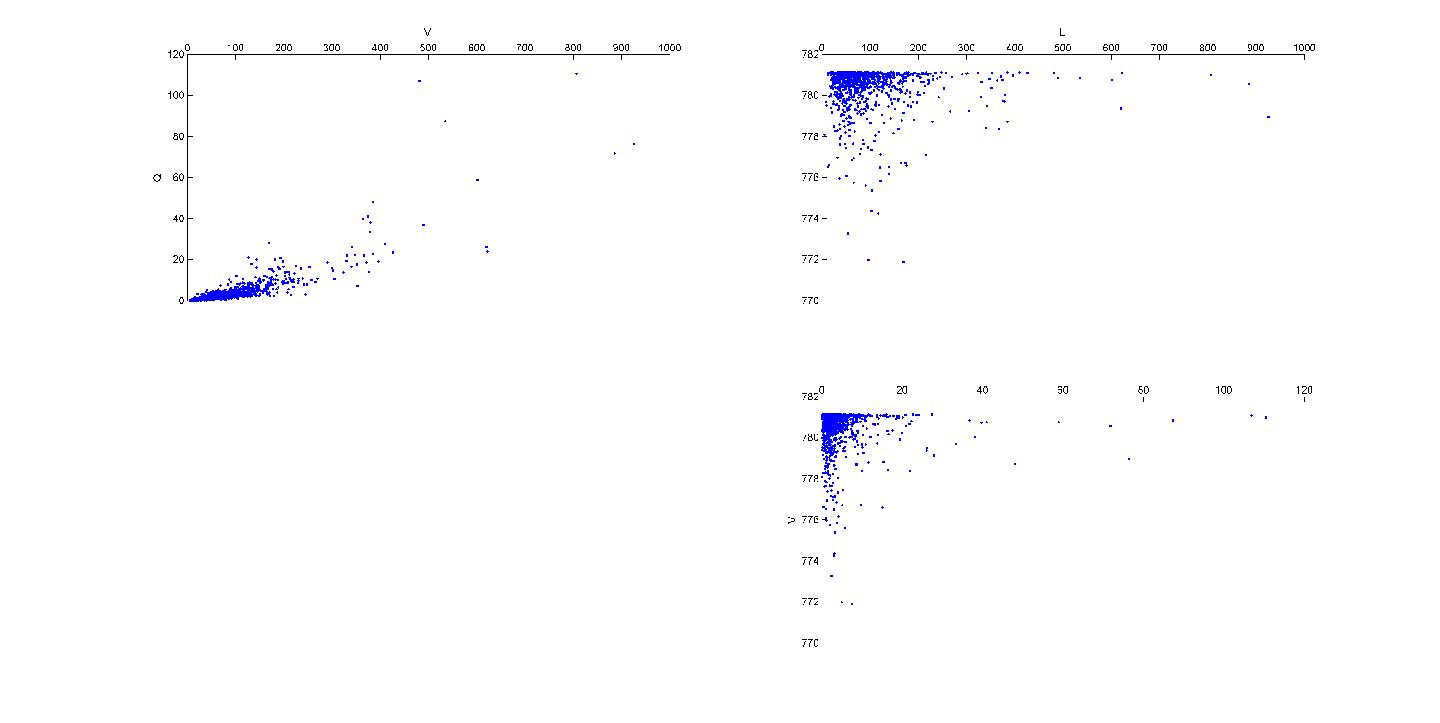}\hspace{-0.5cm}\includegraphics[height=6.5cm,width=6.5cm]{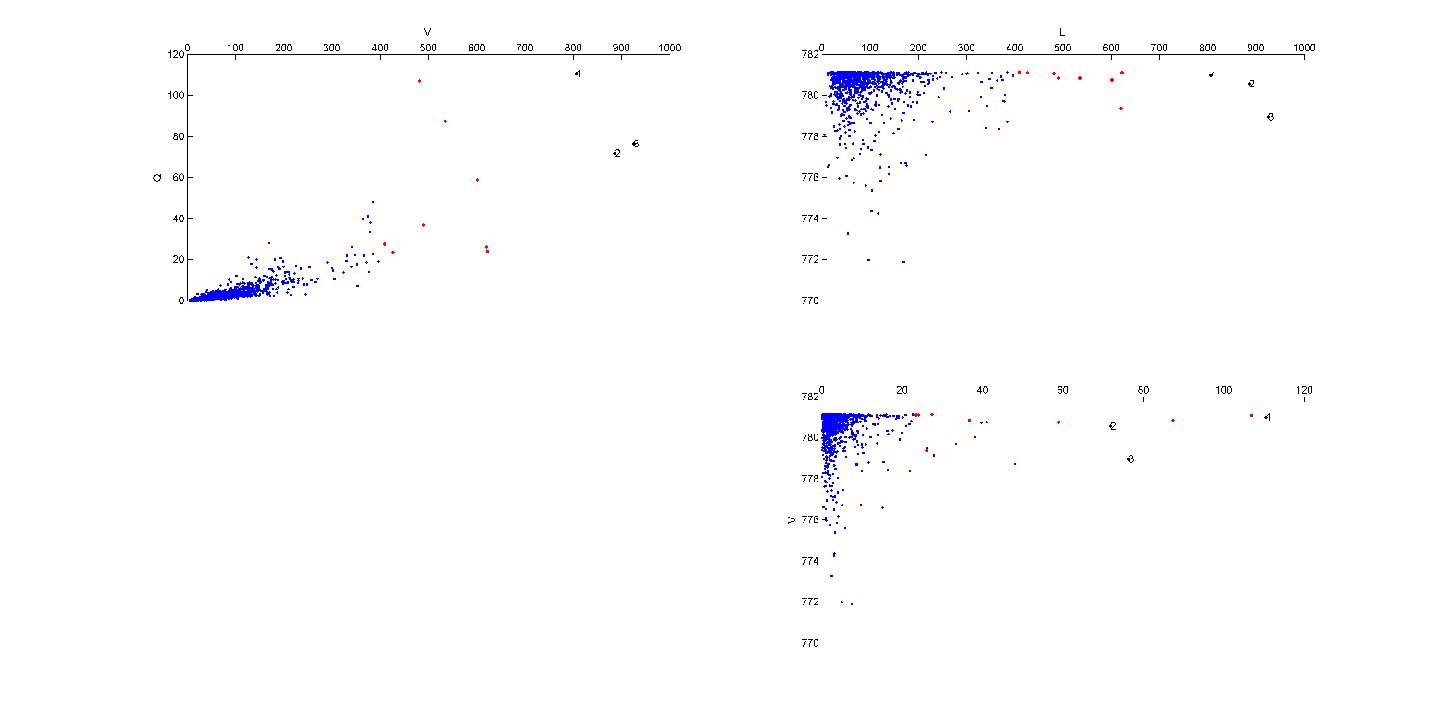}
\centerline{\hspace{1cm}(A)\;$\mathbf{u}=\mathbf{e}$\hspace{4cm} (B)\;$\mathbf{u}=$First \textit{PCA}}\ \ \\
\caption{Directional Extremes through distributions at $99\%$, ($\alpha=1\%$)}\label{fig:simCeppo_eypca_App}
\vspace{-0.5cm}
\end{center}
\end{figure}

The analysis through distributions at level $99\%$ for Ceppo Morelli dam model can be performed through simulation as we did in the paper.  Figure \ref{fig:simCeppo_eypca_App} is an example of the result for the classical and the first \textit{PCA} directions. Notice that Figure \ref{fig:simCeppo_eypca_App} (A) displays the results in the classical direction $\mathbf{e}$, which is an empty identification of extremes. On the other hand, the improvement of using the first \textit{PCA} direction is evident. 

Now, using the stress and reliability method over the behavior of the reservoir routing in horizons of $1000-$years long, that we have proposed in the paper, it is possible to highlight again the advantages of the directional approach. Table \ref{table:extResults_App} summarizes average indexes from the analysis through distribution functions at level $99\%$ over $100$ simulated samples in both directions. Specifically, the table describes: 1) The false positive ratio, which is the number of observations bad identified as critical over the total number of critical identifications. 2) The true positive ratio, which is the number of critical values correctly identified over the total number of real critical values  from the dam routing simulation. 3) The extremes detection ratio, which is the number of observations identified as critical over the total number of observations. 4) The true extremes ratio, which is the number of real critical values over the total number of observations.

\begin{table}[htbp]
\begin{center}
\begin{tabular}{|| l |c|c||}\hline\hline
 & Classical Direction & PCA Direction \\ \hline\hline
False Positives Ratio & $NaN$ & $34.93\%$  \\ \hline
True Positives Ratio &  $0.0\%$ & $99.44\%$ \\ \hline
Extremes Detection Ratio & $0.01\%$ & $1.27\%$  \\ \hline
True Extremes Ratio & $0.83\%$ & $0.83\%$  \\ \hline\hline
\end{tabular}
\end{center}
\caption{\normalsize{Results of the Directional Extreme Analysis through distributions at $99\%$}}\label{table:extResults_App}
\end{table}

The table shows better indexes using the first \textit{PCA} direction than the classical. Note that results between \textit{upper sets} in both approaches are consistent according to equation (2.5) in the paper.

\subsection{Case study: Sea storms}

\begin{figure}[htbp]
\begin{center}
\includegraphics[height=6.5cm,width=6.5cm]{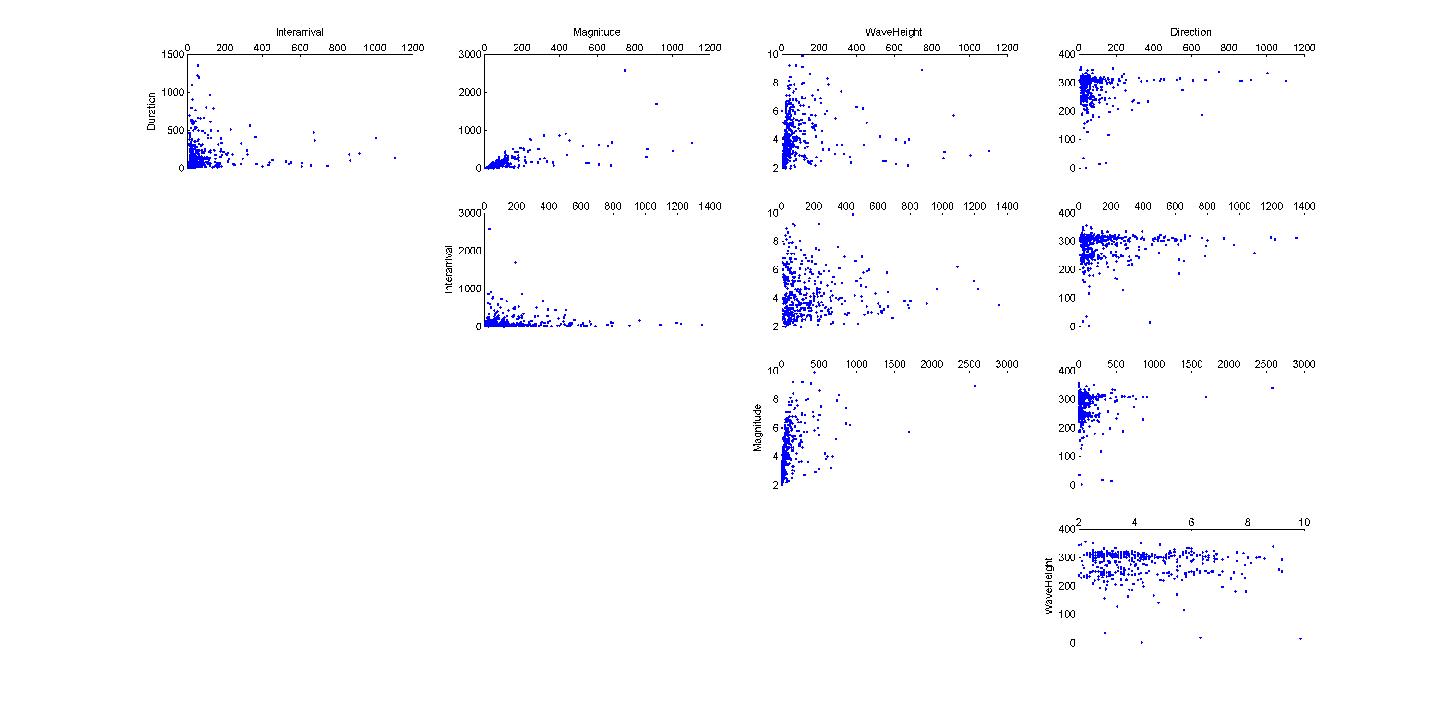}\hspace{-0.5cm}\includegraphics[height=6.5cm,width=6.5cm]{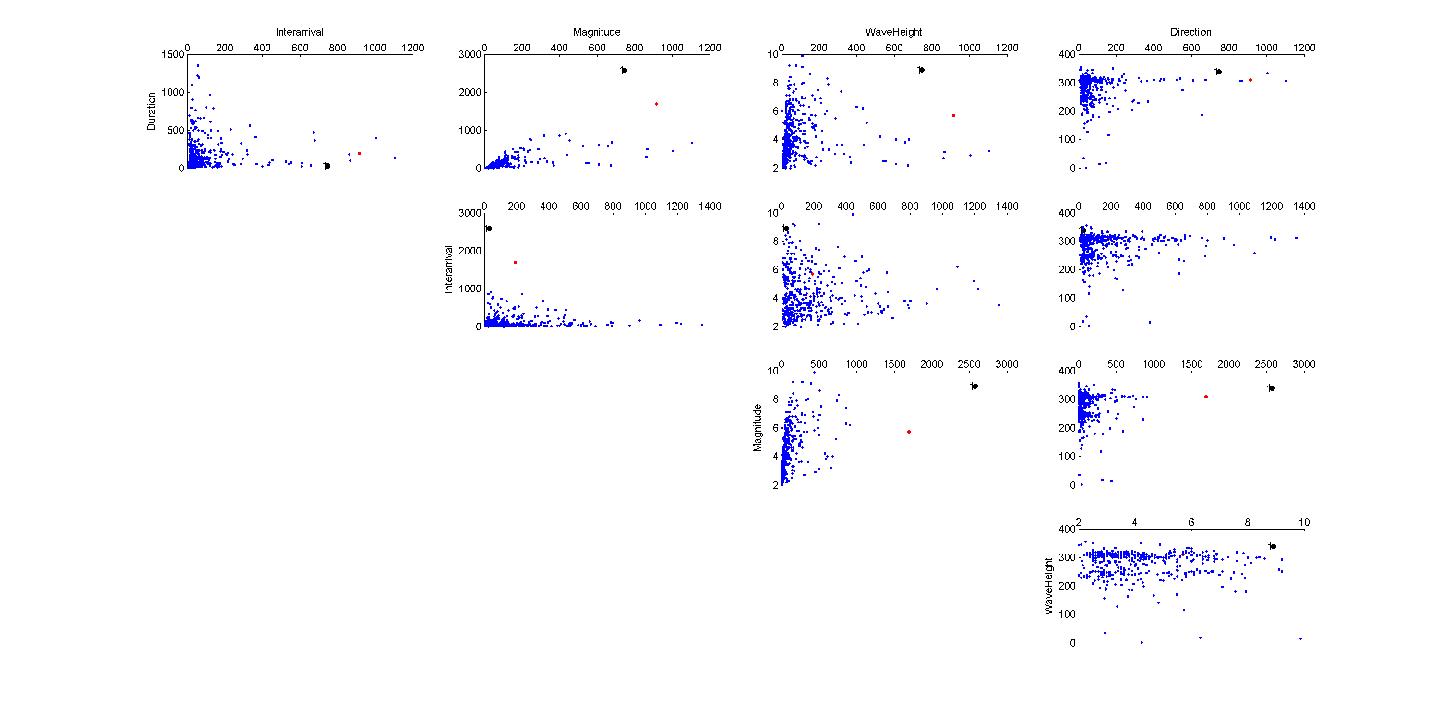}
\centerline{\hspace{1cm}(A)\;$\mathbf{u}=\mathbf{e}$\hspace{4cm} (B)\;$\mathbf{u}=$First \textit{PCA}}\ \ \\
\caption{Directional Extremes throug distributions at level $99\%$, ($\alpha=1\%$)}\label{fig:alg_eypca_App}
\vspace{-0.5cm}
\end{center}
\end{figure}

If we analyze the sea storms dataset collected at Alghero, Italy. Figure \ref{fig:alg_eypca_App} (A) presents the results of the classical approach through distributions at level $99\%$ and Figure \ref{fig:alg_eypca_App} (B) displays the results in the first \textit{PCA} direction. The results are similar to those in the flood risk analysis at a dam showing better performance using the directional approach over the first \textit{PCA} direction. 

\subsection{Extremes based on copulas and the directional approach}

Finally, in this section we recall the copula examples used in the paper to show the extreme detection through distribution functions. Specifically, we consider:  1. The elliptical copula example, 2. The positive dependent Frank copula, and 3. The negative dependent Frank copula.

\begin{figure}[htbp]
\begin{center}
\includegraphics[height=30mm,width=30mm]{scatNorm.jpg}\hspace{1cm}
\includegraphics[height=30mm,width=30mm]{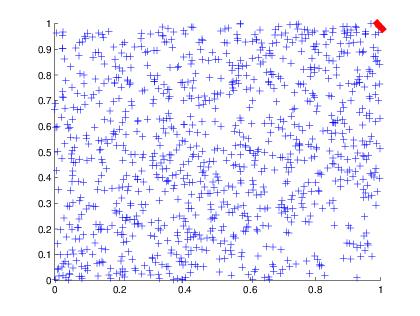}\hspace{1cm}\includegraphics[height=30mm,width=30mm]{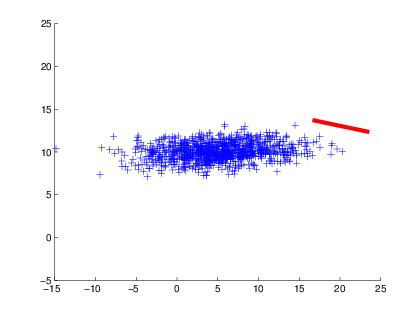}\ \ \\
\centerline{(A)\;Original data\hspace{1.5cm} (B)\;Original data in \hspace{1.5cm} (C)\;Original data}\ \ \\
\vspace{-0.5cm}
\centerline{the copula space}\ \ \\
\includegraphics[height=30mm,width=30mm]{scatNormRot.jpg}\hspace{1cm}
\includegraphics[height=30mm,width=30mm]{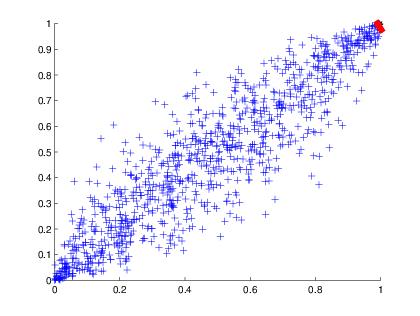}\hspace{1cm}\includegraphics[height=30mm,width=30mm]{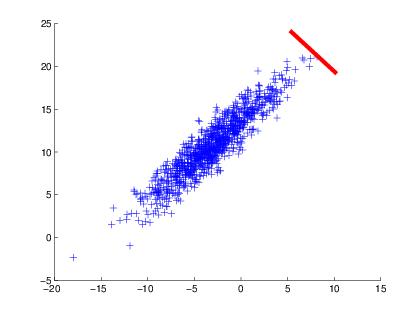}\ \ \\
\centerline{(D)\;Rotated data\hspace{1.2cm} (E)\;Rotated data in\hspace{1.2cm} (F)\;Rotated data}\ \ \\
\vspace{-0.5cm}
\centerline{the copula space}\ \ \\
\caption{Top: theoretical results through the classical distribution approach; Bottom: theoretical results through the classical distribution approach for the rotation of the data in the first \textit{PCA} direction}\label{fig:normCase_App}
\vspace{-0.5cm}
\end{center}
\end{figure}

\begin{figure}[htbp]
\begin{center}
\includegraphics[height=40mm,width=40mm]{normData_-e_99.jpg}\hspace{1cm}
\includegraphics[height=40mm,width=40mm]{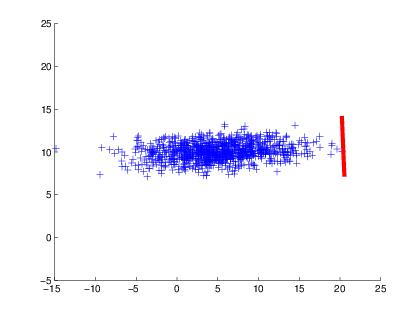}\ \ \\
\centerline{\hspace{0.2cm}(A)\; Classical direction $\mathbf{e}$\hspace{1.3cm} (B)\; First \textit{PCA} direction}\ \ \\
\caption{Comparison of the identification of extremes in the directions $\mathbf{e}$ and first \textit{PCA} (black points)}\label{fig:normResults_App}
\vspace{-0.5cm}
\end{center}
\end{figure}

Firstly, we recall the Gaussian copula example with parameters $\mu_{1} = 5$, $\sigma_{1}^{2}=25$, $\mu_{2} = 10$, $\sigma_{2}^{2}=1$ and $\rho = 0.2$. Figure \ref{fig:normCase_App} summarizes the results of the extreme detection through distribution function approach at level $99\%$ ($\alpha=1\%$). The three top plots describe the procedure in the classical direction for the original data,  meanwhile the three bottom plots describe the results also in the classical direction, but for the data rotated according to the corresponding $R_{\mathbf{u}}$ when $\mathbf{u}$ is the first \textit{PCA} direction.  The quantiles (red) in the copula space are located in the upper-right corner, very close to $(1,1)$, and they can be calculated theoretically in both cases thanks to the elliptical properties. Figure \ref{fig:normResults_App} shows the comparison of the results by undoing the previous rotation (right plot), where can be observed  an improvement in the identification of extremes according to the shape of the data.

Now, we describe the results in a model considering \textit{Frank} copula with dependence parameter $\theta=5$, and marginals given by \textit{GEV} distributions with parameters $\beta_{1}=5$, $\epsilon_{1}=10$, $\gamma_{1}=-1/2$, $\beta_{2}=1/2$, $\epsilon_{2}=2$ and $\gamma_{2}=1$.

\begin{figure}[htbp]
\begin{center}
\includegraphics[height=30mm,width=30mm]{scatDepPos.jpg}\hspace{1cm}
\includegraphics[height=30mm,width=30mm]{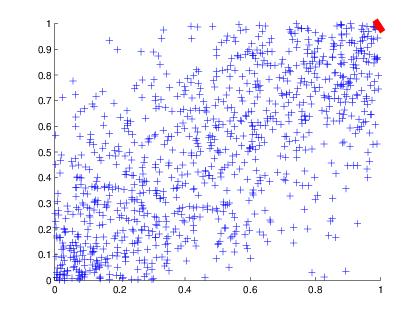}\hspace{1cm}\includegraphics[height=30mm,width=30mm]{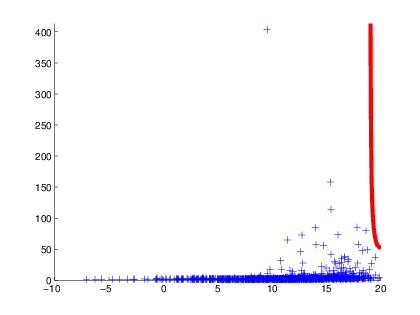}\ \ \\
\centerline{(A)\;Original data\hspace{1.5cm} (B)\;Original data in \hspace{1.5cm} (C)\;Original data}\ \ \\
\vspace{-0.5cm}
\centerline{the copula space}\ \ \\
\includegraphics[height=30mm,width=30mm]{scatDepPosRot.jpg}\hspace{1cm}
\hspace{4cm}
\includegraphics[height=30mm,width=30mm]{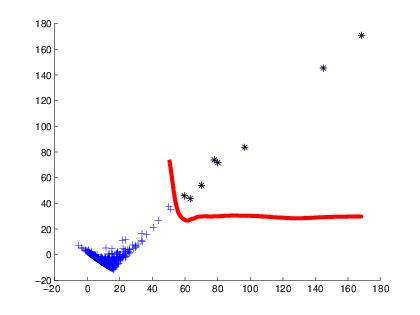}
\centerline{(D)\;Rotated data\hspace{1.2cm} (E)\;Rotated data in\hspace{1.2cm} (F)\;Rotated data}\ \ \\
\centerline{the copula space}\ \ \\
\caption{Top: theoretical results in direction $\mathbf{e}$; Bottom: non-parametric approach in direction $\mathbf{e}$ for the rotation of the data given by the first \textit{PCA} direction}\label{fig:depPos_App}
\vspace{-0.5cm}
\end{center}
\end{figure}

Figure \ref{fig:depPos_App}(A, B, C) show  the classical approach through distributions at level $99\%$ ($\alpha=1\%$) which can be obtained theoretically. However, Figure \ref{fig:depPos_App}(D, E, F) plot the analysis for the same level, but using the non-parametric approach in direction $\mathbf{e}$ over the pre-rotated data under the rotation $R_{\mathbf{u}}$ given by the first \textit{PCA} direction. Figure \ref{fig:depPos_App}(D) shows the data in the rotated space, Figure \ref{fig:depPos_App}(E) is empty due to the absence of theoretical evidence of the copula after the rotation of the data and finally, Figure \ref{fig:depPos_App}(F) indicates the extremes in the rotated space but using the non-parametric approach. 

\begin{figure}[htbp]
\begin{center}
\includegraphics[height=40mm,width=40mm]{depPosData_-e_99.jpg}\hspace{1cm}
\includegraphics[height=40mm,width=40mm]{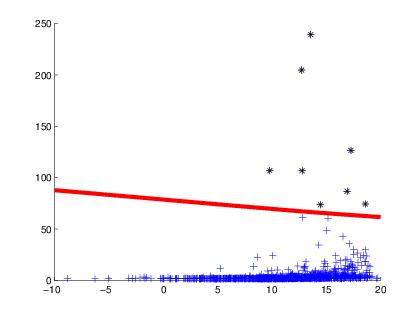}\ \ \\
\centerline{\hspace{0.2cm}(A)\; Classical direction $\mathbf{e}$\hspace{1.3cm} (B)\; First \textit{PCA} direction}\ \ \\
\caption{Comparison of the identification of extremes in the directions $\mathbf{e}$ and first \textit{PCA} (black points)}\label{fig:posResults_App}
\vspace{-0.5cm}
\end{center}
\end{figure}

Figure \ref{fig:posResults_App} displays the results to compare and it is noted that the classical approach through distributions does not identify extremes, even when graphically there exist some of them. On the other hand, the use of the first \textit{PCA} direction in the analysis is more accurate.

\begin{figure}[htbp]
\begin{center}
\includegraphics[height=30mm,width=30mm]{scatDepNeg.jpg}\hspace{1cm}
\includegraphics[height=30mm,width=30mm]{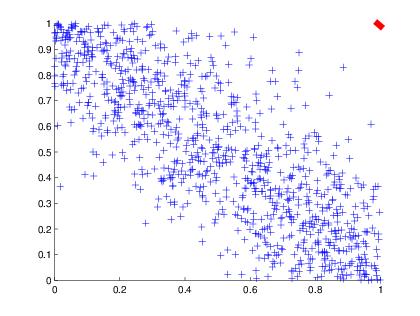}\hspace{1cm}\includegraphics[height=30mm,width=30mm]{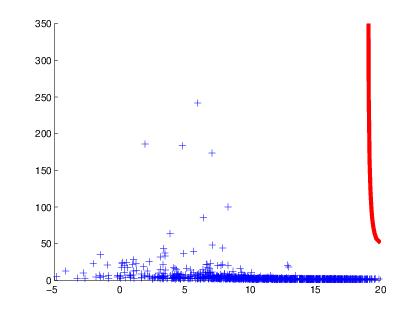}\ \ \\
\centerline{(A)\;Original data\hspace{1.5cm} (B)\;Original data in \hspace{1.5cm} (C)\;Original data}\ \ \\
\vspace{-0.5cm}
\centerline{the copula space}\ \ \\
\includegraphics[height=30mm,width=30mm]{scatDepNegRot.jpg}\hspace{1cm}
\hspace{4cm}
\includegraphics[height=30mm,width=30mm]{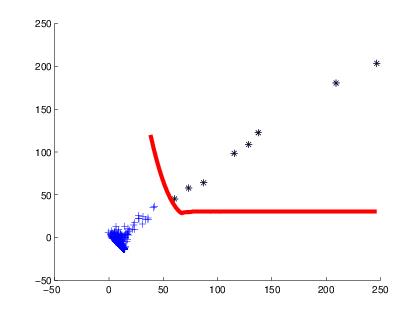}
\centerline{(D)\;Rotated data\hspace{1.2cm} (E)\;Rotated data in\hspace{1.2cm} (F)\;Rotated data}\ \ \\
\centerline{the copula space}\ \ \\
\caption{Top: theoretical results in direction $\mathbf{e}$; Bottom: non-parametric approach in direction $\mathbf{e}$ for the rotation of the data given by the first \textit{PCA} direction}\label{fig:depNeg_App}
\vspace{-0.5cm}
\end{center}
\end{figure}

\begin{figure}[htbp]
\begin{center}
\includegraphics[height=40mm,width=40mm]{depNegData_-e_99.jpg}\hspace{1cm}
\includegraphics[height=40mm,width=40mm]{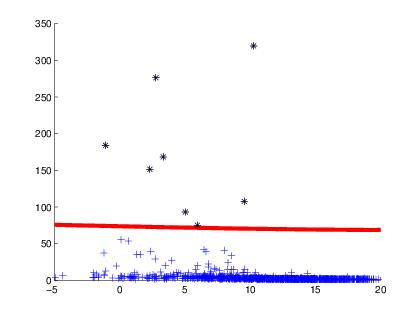}\ \ \\
\centerline{\hspace{0.2cm}(A)\; Classical direction $\mathbf{e}$\hspace{1.3cm} (B)\; First \textit{PCA} direction}\ \ \\
\caption{Comparison of the identification of extremes in the directions $\mathbf{e}$ and first \textit{PCA} (black points)}\label{fig:negResults_App}
\vspace{-0.5cm}
\end{center}
\end{figure}

Finally, we perform the study in the model considering the \textit{Frank} copula with dependence parameter $\theta=-8$, and marginals given by the same \textit{GEV} distributions. Figure \ref{fig:depNeg_App} shows the outputs in the same framework as Figure \ref{fig:depPos_App}, and Figure \ref{fig:negResults_App} shows the contrast between the classical and the first \textit{PCA} directions for the detection of extremes. Once again, we can observe a better pattern of extreme recognition by considering the first \textit{PCA} direction as an alternative in the analysis. 

\end{document}